\documentclass[useAMS,usenatbib]{mn2e}
\usepackage{amsmath}
\usepackage{threeparttable}
\usepackage{multirow}
\usepackage{graphicx}

\def\apj    {{ApJ}}
\def\apjl   {{ApJL}}
\def\apjs   {{ApJS}}
\def\aj     {{AJ}}
\def\aaa    {{A\&A}}

\def\araa   {{ARA\&A}}

\def\mnras  {{MNRAS}}

\def\sur{\rm mag\,\, arcsec^{-2}}
\def\umag{u^{\prime}}
\def\gmag{g^{\prime}}
\def\rmag{r^{\prime}}

\def\lesssim{\mathrel{\hbox{\rlap{\hbox{\lower4pt\hbox{$\sim$}}}\hbox{$<$}}}}

\title{Impact of Galaxy Mergers on the Colours of Cluster Galaxies}

\author[S.~Oh et al.]
{\parbox{\textwidth}{Sree~Oh$^{1,2,3}$\thanks{E-mail: sree.oh@anu.edu.au},
Keunho Kim$^{4}$,
Joon Hyeop Lee$^{5, 6}$,
Minjin Kim$^{5,7}$,
Yun-Kyeong Sheen$^{5}$,
Jinsu Rhee$^{3}$,
Chang H. Ree$^{5}$,
Hyunjin Jeong$^{5}$,
Luis C. Ho$^{8,9}$,
Jaemann Kyeong$^{5}$,
Eon-Chang Sung$^{5}$,
Byeong-Gon Park$^{5, 6}$,
Sukyoung K. Yi$^{3}$} \vspace{0.4cm}\\
\parbox{\textwidth}{$^{1}$ Research School of Astronomy and Astrophysics, Australian National University, Canberra, ACT 2611, Australia\\
$^{2}$ ARC Centre of Excellence for All Sky Astrophysics in 3 Dimensions (ASTRO 3D), Australia\\
$^{3}$ Department of Astronomy \& Yonsei University Observatory, Yonsei University, Seoul 03722, Republic of Korea\\
$^4$ School of Earth \& Space Exploration, Arizona State University, Tempe, AZ 85287, USA \\
$^5$ Korea Astronomy and Space Science Institute, Daejeon 34055, Republic of Korea\\
$^6$ University of Science and Technology, Daejeon 34113, Republic of Korea\\
$^7$ Department of Astronomy and Atmospheric Sciences, College of Natural Sciences, Kyungpook National University, Daegu 41566, Republic of Korea\\
$^8$ Kavli Institute for Astronomy and Astrophysics, Peking University, Beijing 100871, China\\
$^9$ Department of Astronomy, School of Physics, Peking University, Beijing 100871, China}}

\begin{document}
\date{Accepted 00. Received 00; in original form 00}

\pagerange{\pageref{firstpage}--\pageref{lastpage}} \pubyear{2018}
\maketitle
\label{firstpage}

\begin{abstract}
We examine the ultraviolet and optical colours of 906 cluster galaxies from the KASI-Yonsei Deep Imaging Survey of Clusters (KYDISC). Galaxies have been divided into two categories, morphologically-disturbed and undisturbed galaxies, based on the visual signatures related to recent mergers. We find that galaxies with signatures of recent mergers show significantly bluer colours than undisturbed galaxies. Disturbed galaxies populate more on the cluster outskirts, suggesting recent accretion into the cluster environment, which implies that disturbed galaxies can be less influenced by the environmental quenching process and remain blue. However, we still detect bluer colours of disturbed galaxies in all locations (cluster core and outskirts) for the fixed morphology, which is difficult to understand just considering the difference in time since infall into a cluster. Moreover, blue disturbed galaxies show features seemingly related to recent star formation. Therefore, we suspect that mergers make disturbed galaxies keep their blue colour longer than undisturbed galaxies under the effect of the environmental quenching through either merger-induced star formation or central gas concentration which is less vulnerable for gas stripping.
\end{abstract}

\begin{keywords}
galaxies: clusters: general -- galaxies: interactions -- galaxies: fundamental parameters -- galaxies: evolution
\end{keywords}

\section{Introduction}
\label{sec:intro}
  The colour-magnitude relation mirrors stellar masses and populations of galaxies, two key characteristics of galaxies, and therefore is conventionally used to differentiate galaxy types (e.g. Baldry et al. 2004; Wyder et al. 2007). Galaxies have often been classified into three main groups based on their location on the colour-magnitude diagram (CMD): the red sequence which is dominated by early-type and/or passively evolving populations, the blue cloud where late-type and/or actively star-forming galaxies mainly reside, and the green-valley where we can find the mixing of morphology and intermediate star-formation rate. In general, we expect current star-forming and passively evolving populations in the blue cloud and red sequence respectively (Renzini 2006).

 Since the Galaxy Evolution Explorer (GALEX; Martin et al. 2005) data became available, ultraviolet (UV) CMDs have been utilised to identify the recent episode of star-formation. The significant fraction of residual star formation in early-type galaxies was first introduced in Yi et al. (2005), and many studies after reported that 10--15\% of bright early-type galaxies show signatures of recent star formation based on the UV CMD (e.g., Kaviraj et al. 2007; Jeong et al. 2007, 2009). Even in optical CMDs, blue early-type galaxies have been reported, suggesting young stellar populations in early-type galaxies (Lee, Lee, \& Hwang 2006; Schawinski et al. 2007, 2009; McIntosh et al. 2014). 
 
 Galaxy mergers are considered one of the primary sources for blue early-type galaxies (George \& Zingade 2015; Haines et al. 2015; George 2017). In the past decade, many observational and theoretical studies have supported merger-induced star formation (e.g. Sanders et al. 1988; Lawrence et al. 1989; Mihos \& Hernquist 1996; Duc et al. 1997; Wang et al. 2004; Cox et al. 2008; Di Matteo et al. 2008). Young stars generated by mergers would bring a blue colour in galaxies, which suggests a migration on CMDs.
 
 Recent studies using deep photometry revealed a considerable fraction ($\sim$20\%) of cluster galaxies showing signatures of post- and ongoing mergers (Sheen et al. 2012; Oh et al. 2018). In addition, Sheen et al. (2016) found that cluster galaxies with post-merger signatures which were visually identified from deep images show a higher fraction of UV blue galaxies compared to normal galaxies that reside in the optical red sequence. Therefore, galaxy interactions are suspected of having a close relation to recent star formation in galaxies, even in the cluster environment.
 
 However, the investigations on the effect of mergers based on CMDs were biased to early-type galaxies perhaps because of difficulty in assessing merger-driven changes in late-type galaxies which naturally show ongoing star formation. Studies on CMDs in the cluster environments have also focused on the passive population and the possible mechanisms causing a migration toward the red sequence because the majority of cluster populations in the nearby universe are passive early-type galaxies especially within the cluster virial radius (e.g. Blakeslee et al. 2003; L$\acute{\rm o}$pez-Cruz et al. 2004; Mei et al. 2006, 2009).
 
 We investigate the impact of galaxy interactions in optical and UV colours based on the KASI-Yonsei Deep Imaging Survey of Clusters (KYDISC; Oh et al. 2018) catalogue which covers two times virial radius ($R_{200}$) of clusters and allows us to have a considerable number of late-type galaxies in the cluster environment.

Throughout the paper, we adopt Planck 2015 results with $\Omega_{\rm m} = 0.308$, $\Omega_\Lambda= 0.692$, and $H_0 = 67.8$ km/s/Mpc (Ade et al. 2016).


\section{Data and sample}
\label{sec:data}
 
We used a catalogue of 1409 galaxies from the KYDISC targeting 14 clusters at $0.015 \lesssim z \lesssim 0.144$. The KYDISC has used the Inamori Magellan Areal Camera and Spectrograph (IMACS) on the 6.5-meter Magellan Baade telescope and the MegaCam on the 3.6-meter Canada-France-Hawaii Telescope (CFHT) for deep photometry ($\mu_{\rmag}\sim$ 27 $\sur$) and provides a catalogue including $\umag$, $\gmag$, and $\rmag$ magnitudes, redshifts, morphology, bulge-to-total ratio (B/T), and local density.

\begin{figure}
      \centering
       \includegraphics[width=0.5\textwidth]{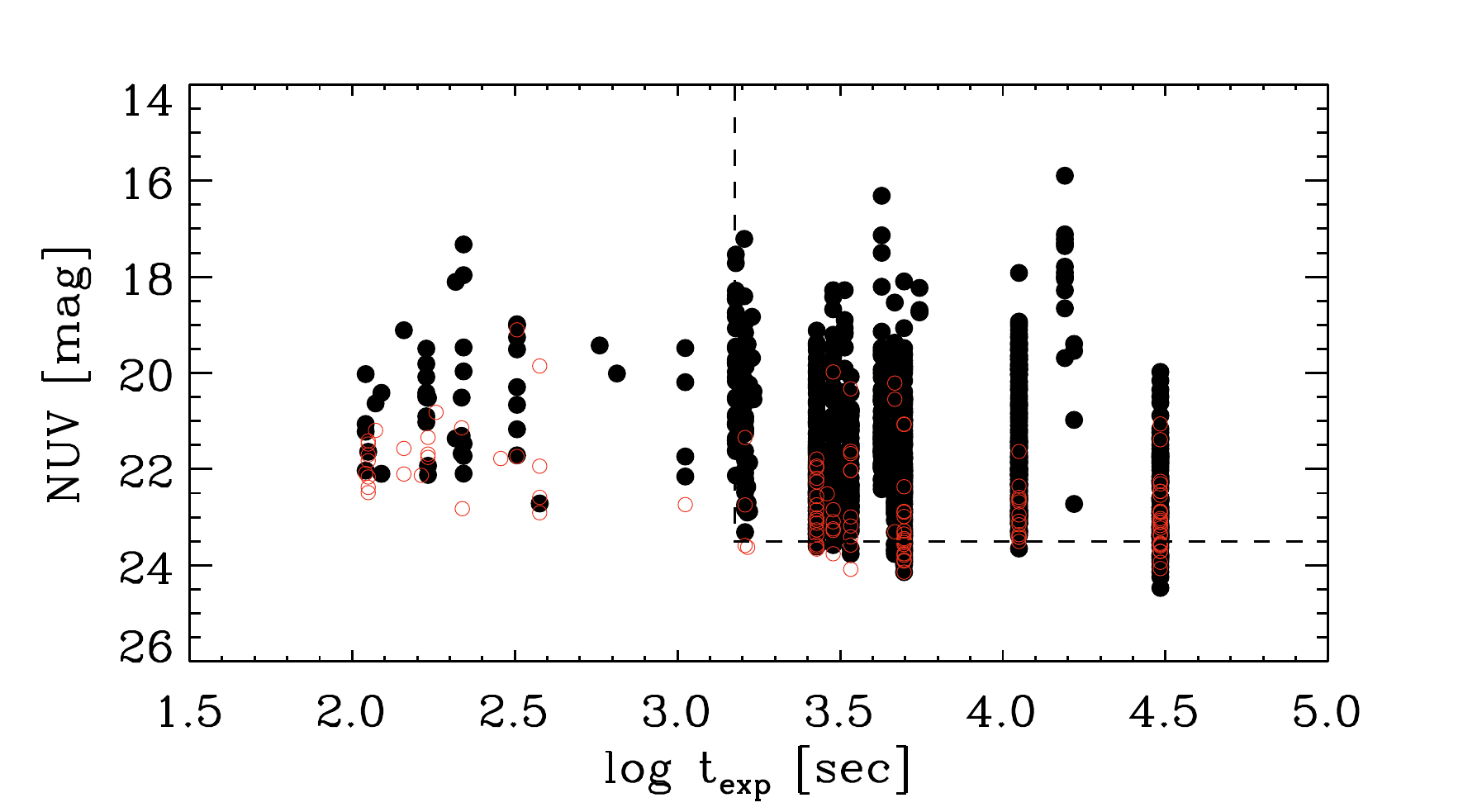}
       \caption{The NUV exposure time and magnitude for 1027 galaxies with NUV detections. Black-filled circles are 906 sample galaxies, and red-open circles show the 121 galaxies excluded from the sample after considering aperture sizes and magnitude errors in the NUV band. The impact of the conservative sampling (dashed line; NUV $<$ 23.5 and $t_{\rm exp}>1500$ seconds) on the main result is not significant (Appendix). }
       \label{sample}
     \end{figure}

Near-ultraviolet (NUV; $\lambda_{\rm eff}=2267\AA$) flux was obtained from the GALEX GR 6 plus 7 Data Release. We match the cluster galaxies from the KYDISC to GALEX detections within 4$''$, the spatial resolution of GALEX, and have 1027 matches with NUV detections. We have excluded 87 galaxies that are 3$\sigma$ outliers in the relation of aperture size ratio between NUV and $\rmag$ bands to the NUV magnitude in order to reduce mismatches and contaminations by nearby sources. Additional 34 galaxies with magnitude error in NUV greater than 0.5 have been excluded. Finally, we have 906 cluster galaxies having both optical and NUV data. We included NUV detections from four separate GALEX surveys where NUV exposure times are varying from 100 to 30,000 seconds; however, 95\% of the sample have an NUV exposure time greater than 1500 seconds (Figure~\ref{sample}). We present the result from the conservative sampling (NUV $<$ 23.5 and $t_{\rm exp}>1500$) in the Appendix. The optical and NUV magnitudes were corrected for the Galactic extinction using dust maps from Schlafly \& Finkbeiner (2011). We also carried out $k$-correction following the method described in Chilingarian, Melchior, \& Zolotukhin (2010).

\begin{table}
\centering
\caption[Summary of the classification]{Summary of the classification}
\begin{tabular}{l c c c c c c} \hline \hline
& E & S0 & S$_{\rm E}$ & S$_{\rm L}$ & Total\\
\hline
Undisturbed & 237 & 252 & 97 & 67 & 653\\
Disturbed & 34 & 85 & 87 & 47 & 253\\
\hline
Total & 271 & 337 & 184 & 114 & 906\\
\hline
\hline
\end{tabular}
\label{tab:class}
\end{table} 

\begin{figure}
      \centering
       \includegraphics[width=0.5\textwidth]{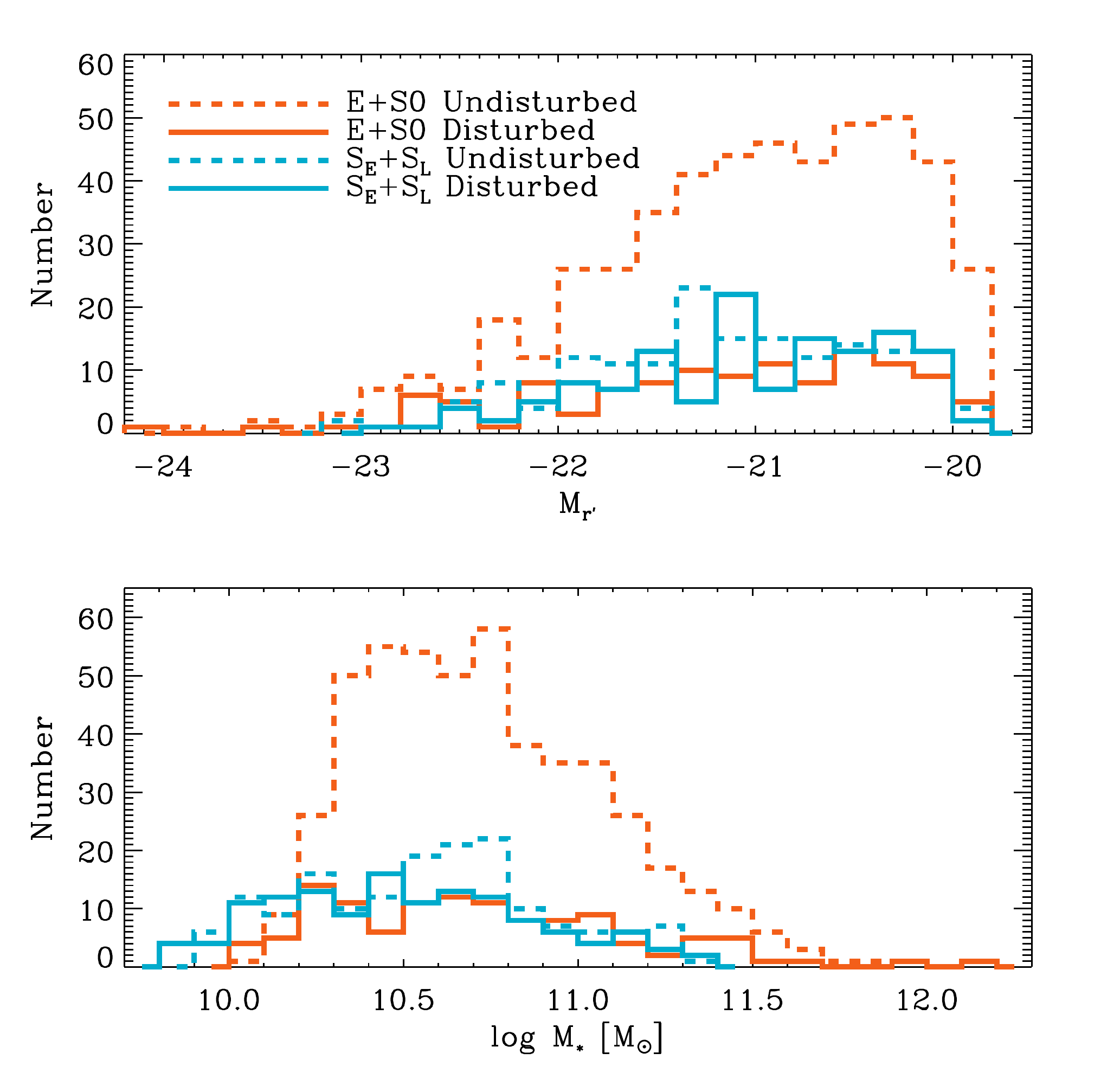}
       \caption{The distributions of the absolute $\rmag$ magnitude (top) and mass (bottom) for 906 sample galaxies. Disturbed and undisturbed groups show similar distributions in magnitude and mass.}
       \label{mag}
     \end{figure}
     
Following the Oh et al. (2018) classification on morphology, the sample galaxies have been classified into four groups: ellipticals (E), lenticulars (S0), early spirals (S$_{\rm E}$), and late spirals (S$_{\rm L}$). The KYDISC catalogue also identified galaxies with signatures of mergers, relying on visual disturbances in deep images. The classification has been summarised in Table~\ref{tab:class}. We adopted the classification and divided galaxies into two groups: disturbed galaxies showing signatures of post or ongoing mergers (PM and OM from Oh et al. 2018) and undisturbed galaxies without any features related to galaxy interactions (N from Oh et al. 2018). Note that both Oh et al. (2018) and this study assume that merger events that made the features in disturbed galaxies have preceded the galaxy accretion into the cluster environment. We found comparable distributions in magnitude and stellar mass between disturbed and undisturbed galaxies for 906 sample galaxies (Figure~\ref{mag}). We have used a B/T from the KYDISC catalogue where galaxies have been decomposed into free S\'{e}rsic (bulge) and exponential (disk) components based on  $\rmag$-band images using {\sc{galfit}} software (Peng et al. 2002, 2010). For more information on the visual classifications and parameters used in this study, refer to Oh et al. (2018).

\begin{figure}
       \centering
       \includegraphics[width=0.5\textwidth]{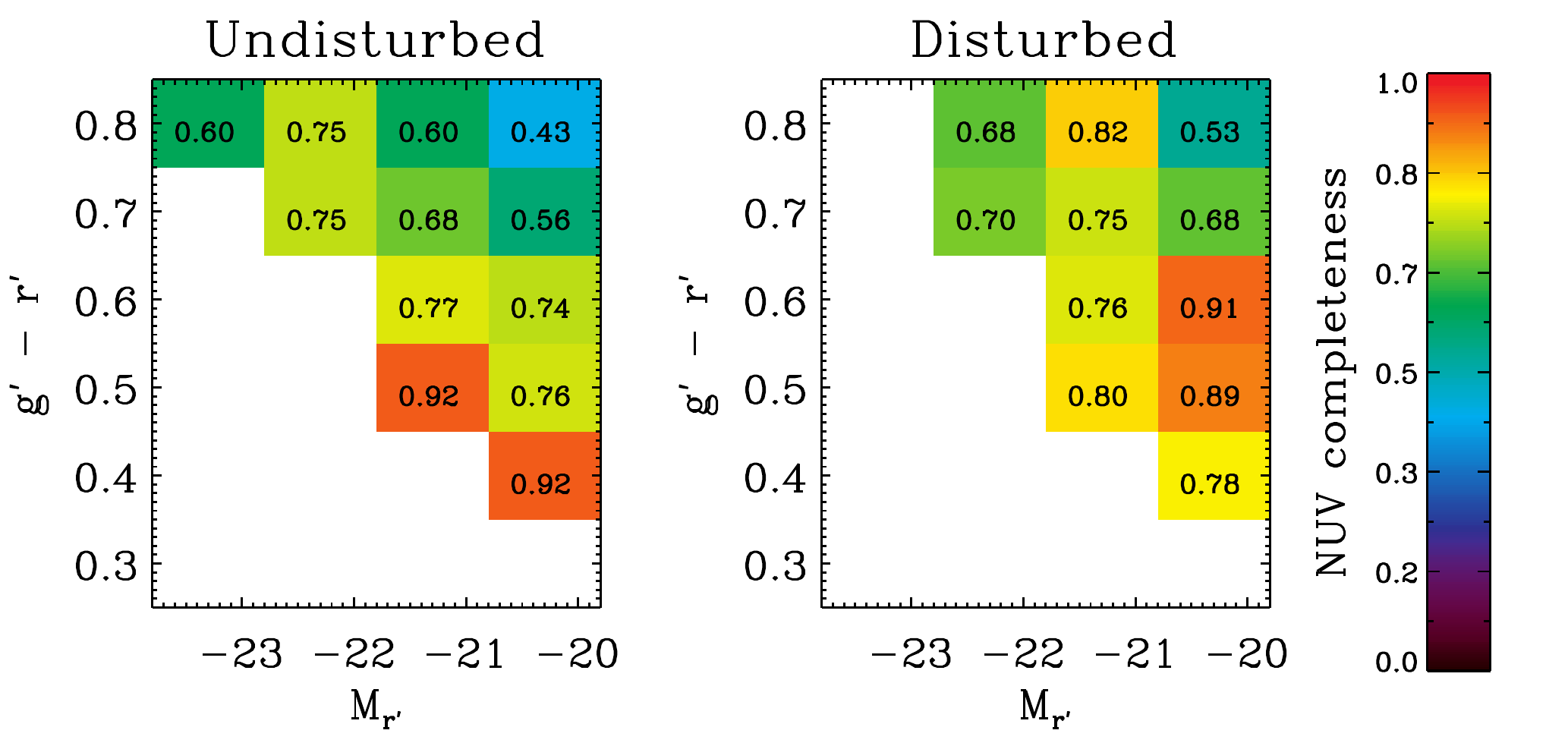}
       \caption[The NUV exposure time and magnitude error]
      {The NUV completeness for undisturbed (left) and disturbed galaxies (right). The reciprocal of the NUV completeness is used for the number weights throughout this study. }
       \label{galex}
     \end{figure}

Our sample based on the NUV detection is biased toward galaxies with young stellar populations (Figure~\ref{galex}). Oh et al. (2018) reported that merger-related galaxies (disturbed galaxies in this study) have bluer NUV$-\rmag$ colours than galaxies without merger signatures. As a result, our sample displays a higher fraction of disturbed galaxies (28$\pm1$\%) than the KYDISC program (24$\pm1$\%). We found a little difference between disturbed and undisturbed galaxies in the NUV completeness for the same colour-magnitude bin which might affect our results mainly when comparing disturbed and undisturbed galaxies. Throughout the paper, we have accounted for the bias embedded in our sampling by applying number weights, the reciprocal of the NUV completeness according to $\gmag-\rmag$ and M$_{\rmag}$. However, we do not find qualitative differences in our results when number weights are not applied.

\section{The optical colour-magnitude relation}
\label{sec:gr}
\begin{figure}
       \centering
       \includegraphics[width=0.52\textwidth]{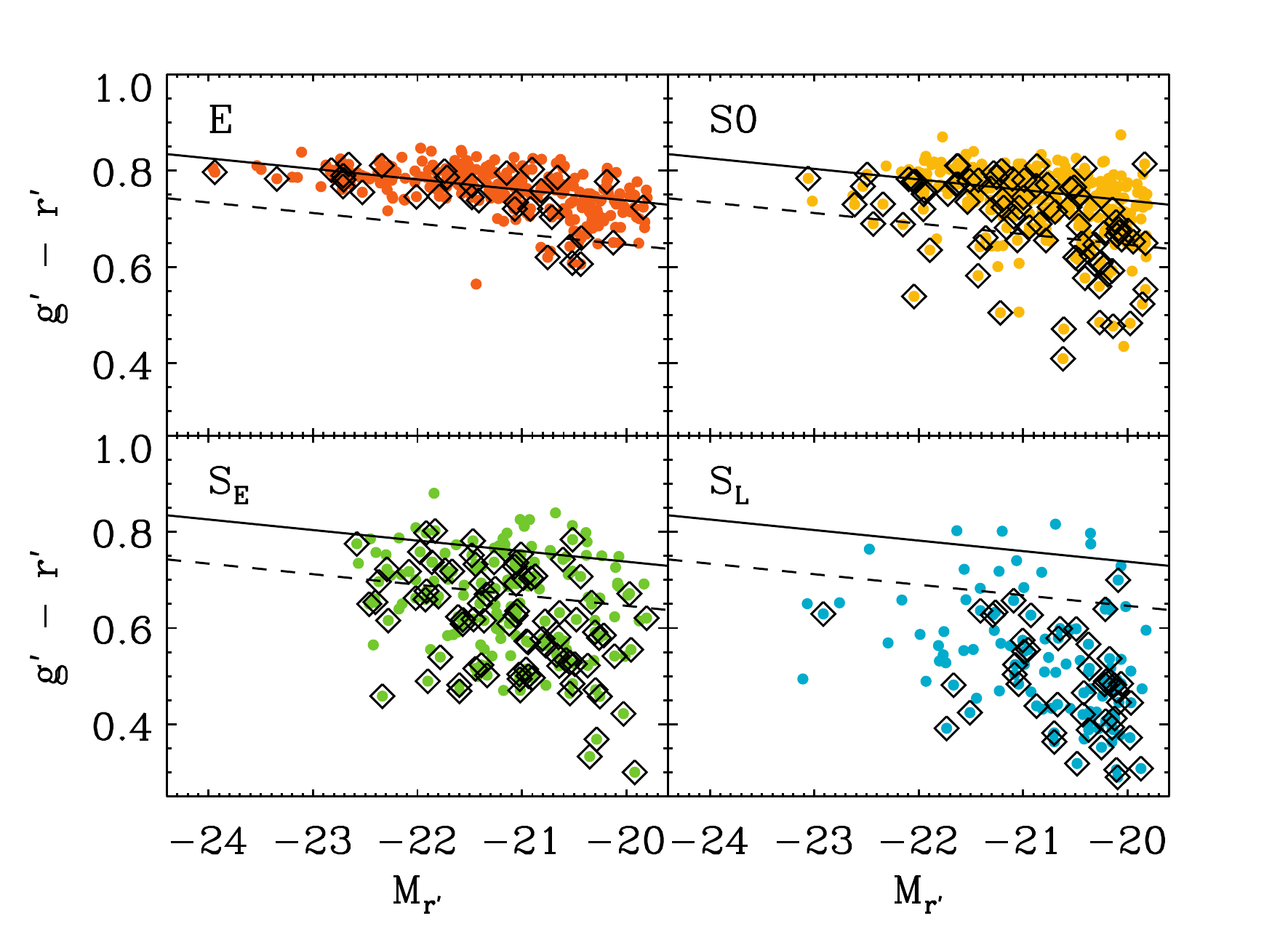}
       \caption[The composite $\gmag-\rmag$ colour-magnitude relation]
      {The composite $\gmag-\rmag$ CMD for each morphology. Solid and dashed lines indicate the RS and the sequence $2\sigma$ below the RS based on elliptical galaxies (E), respectively. Morphologically-disturbed galaxies which are denoted by diamonds show bluer colour compared to undisturbed galaxies in the same morphology.}
     \label{cmdall}
     \end{figure}

\begin{figure}
       \centering
       \includegraphics[width=0.52\textwidth]{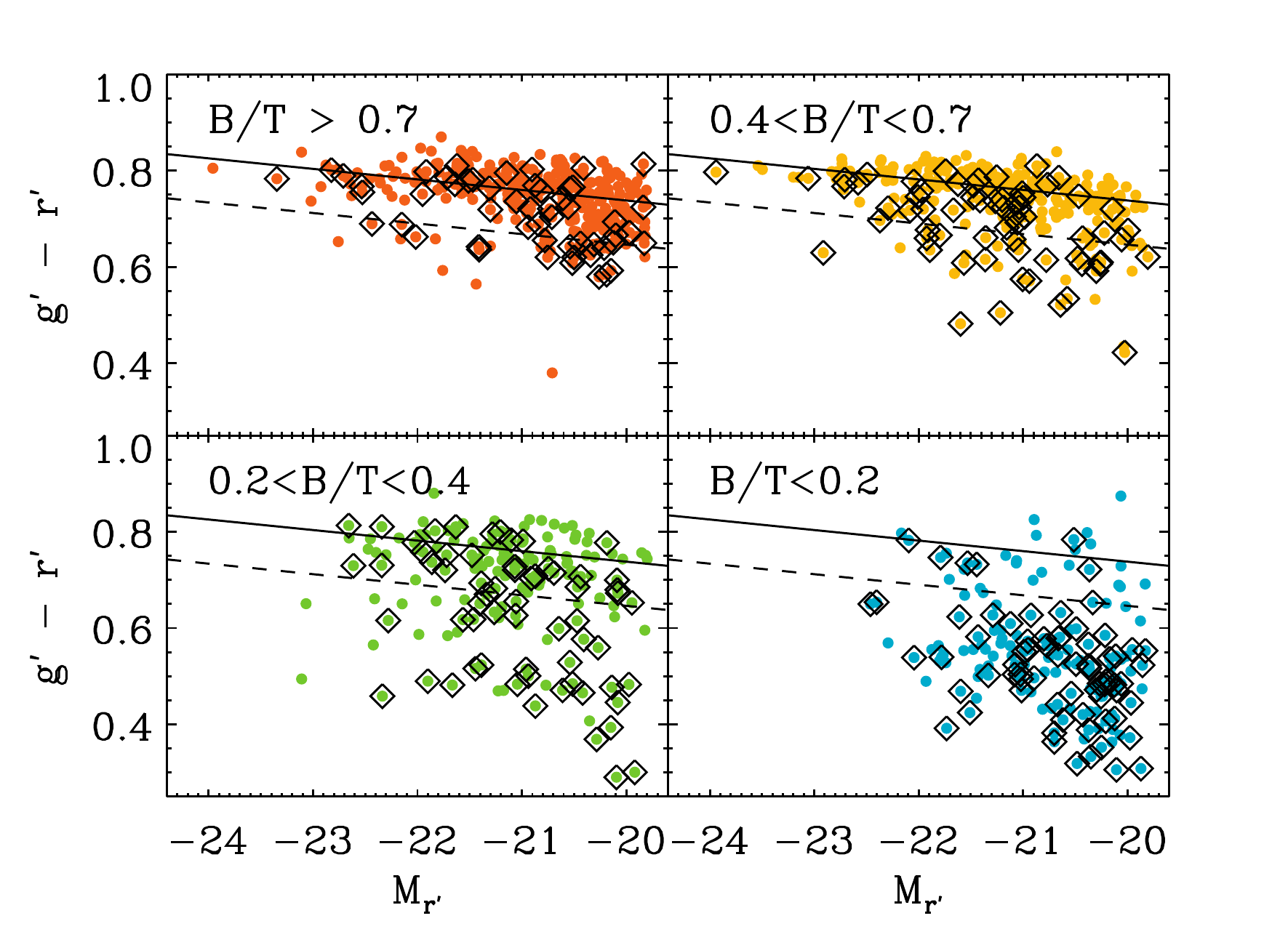}
       \caption[The composite $\gmag-\rmag$ colour-magnitude relation]
      {The composite $\gmag-\rmag$ CMD for each B/T subgroup. Details are the same as Figure~\ref{cmdall}. }
     \label{cmdbt}
     \end{figure}

 We present the optical CMDs for each morphology and B/T in Figures~\ref{cmdall} and \ref{cmdbt}, respectively. A significant fraction of our sample is early-type galaxies (608/906) that mainly populate in the red sequence (RS). We also have 298 late-type galaxies in our sample mostly showing blue $\gmag-\rmag$ colour. Morphologically-disturbed galaxies (diamonds) display a similar range of magnitude with undisturbed galaxies under the same morphology or B/T. However, they exhibit bluer colour than undisturbed galaxies, especially in galaxies with early morphology (E and S0) or high B/T (B/T $>$ 0.4). 

The representative RS on the CMD has been defined based on 271 visually-classified elliptical galaxies. The linear least-squares fit of the elliptical galaxies yields the RS of $\gmag-\rmag\,=\,-0.022\,M_{\rmag}+0.300$. For quantitative comparison, we divided galaxies into three groups based on their location on the CMD: redder red sequence galaxies (RRS; galaxies redder than the RS), bluer red sequence galaxies (BRS; galaxies bluer than the RS and redder than the sequence $2\sigma$ below from the RS), and blue galaxies (BG; galaxies bluer than the sequence $2\sigma$ below from the RS). 

\begin{figure}
       \centering
       \includegraphics[width=0.5\textwidth]{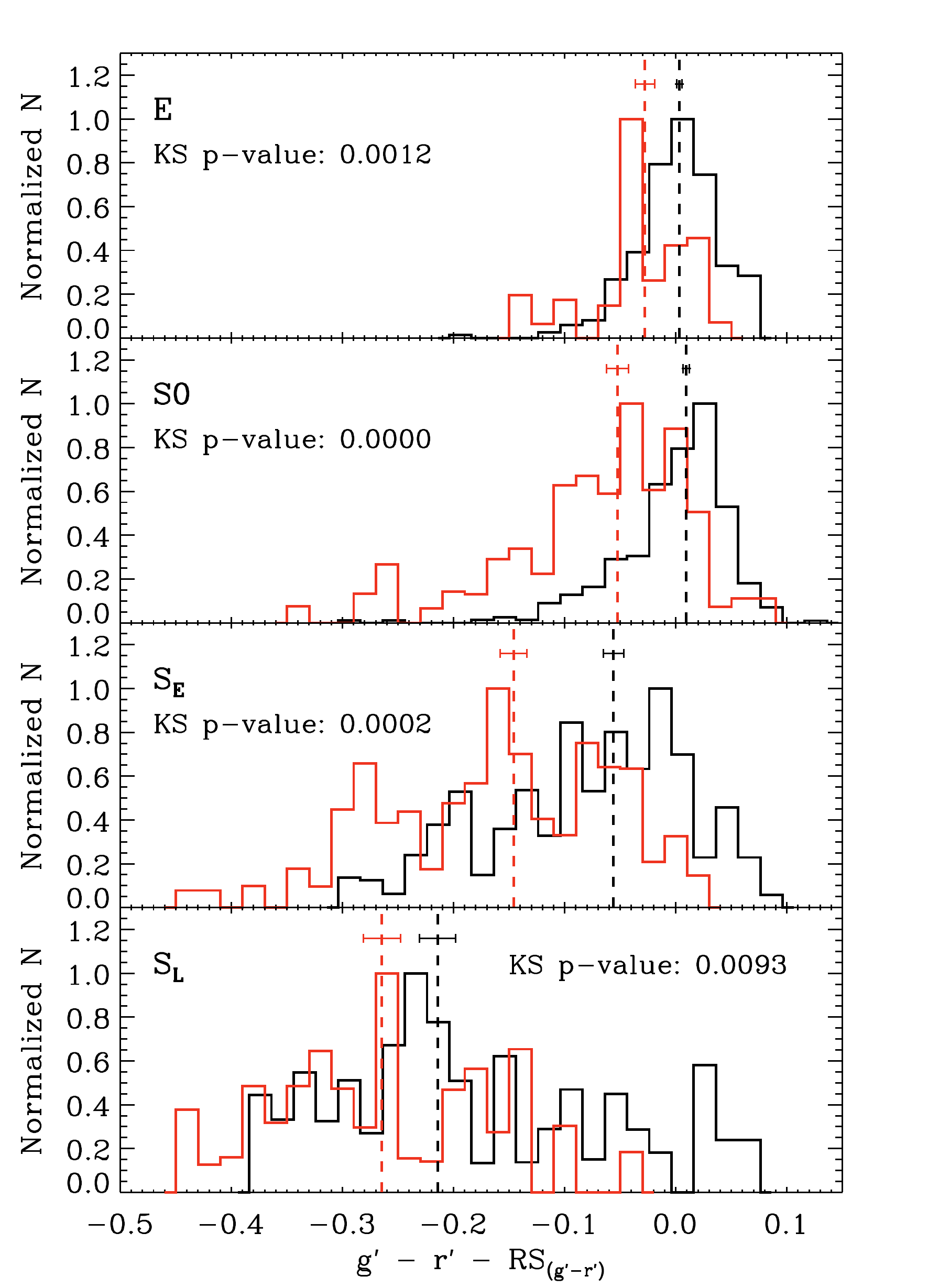}
       \caption[The $\gmag-\rmag$ colour distribution]
      {The $\gmag-\rmag$ colour distribution corrected for the NUV completeness. Dashed lines indicate median value of colour, suggesting bluer colour distribution of disturbed galaxies (red) compared to undisturbed ones (black). Error bars represent standard error of the median. The median p-value from the thousand iterations of the KS test considering $\gmag-\rmag$ errors is shown in each panel.}
     \label{cmddis1}
     \end{figure}

The statistics for each morphology suggest that morphologically-disturbed galaxies have a bluer optical colour than their counterparts in all morphologies (Table~\ref{tab:optcol}). Undisturbed elliptical galaxies are equally distributed between RRS and BRS, whereas disturbed ones mainly belong to BRS. Lenticular galaxies also show a similar result with ellipticals: most undisturbed galaxies belong to RS groups, while disturbed lenticulars are lying in BRS and  BG regions on the CMD. As a result, disturbed galaxies show a five-times-higher fraction of BG (27$\pm7$\%, 32/119) than undisturbed galaxies (5$\pm2$\%, 24/489) in early-type galaxies (E and S0). We also discovered differing colour distributions between disturbed and undisturbed galaxies in late-type galaxies (S$_{\rm E}$ and S$_{\rm L}$). The colour distribution of disturbed S$_{\rm E}$ and S$_{\rm L}$ seems to be mildly shifted to bluer colours compared to that of undisturbed ones. We found a comparable result based on B/T classification (Figure~\ref{cmdbt}). Figure~\ref{cmddis1} shows the $\gmag-\rmag$ colour distribution between disturbed and undisturbed galaxies for each morphology. The $\gmag-\rmag$ colour was corrected to be the difference between the colour and the RS for simplicity ($\gmag-\rmag-$RS$_{\rm \gmag-\rmag}$). Morphologically-disturbed galaxies indeed populate more on blue $\gmag-\rmag$ colour compared to their counterparts. For quantitative comparison, we measured p-values from the Kolmogorov-Smirnov (KS) test for the thousand sets of colour distributions where galaxy colours have been randomly chosen within errors, and we take the median value for the representative p-value. The extremely low p-values support distinct colour colour distributions in disturbed and undisturbed galaxies.

      \begin{table*}
        \centering
        \caption[The fractions of optical colour groups]{The fractions of optical colour groups}
        \begin{tabular}{lllllll}
        \hline \hline
      Morphology$^{a}$ & Merger   &  N$^{b}$ & RRS$^{c}$ & BRS$^{c}$ & BG$^{c}$\\
        \hline \hline
\multirow{2}{*}{E} & Disturbed &34 (47) & 0.29$\pm$0.13 (0.31$\pm$0.11) &0.59$\pm$0.14 (0.59$\pm$0.12) &0.12$\pm$0.09 (0.09$\pm$0.07) \\
                        & Undisturbed &237 (395) & 0.51$\pm$0.05 (0.54$\pm$0.04) &0.47$\pm$0.05 (0.44$\pm$0.04) &0.02$\pm$0.02 (0.02$\pm$0.01) \\
\hline
\multirow{2}{*}{S0} & Disturbed &85 (115) & 0.19$\pm$0.07 (0.20$\pm$0.06) &0.48$\pm$0.09 (0.50$\pm$0.08) &0.33$\pm$0.08 (0.30$\pm$0.07) \\
                        & Undisturbed &252 (445) & 0.54$\pm$0.05 (0.58$\pm$0.04) &0.38$\pm$0.05 (0.36$\pm$0.04) &0.08$\pm$0.03 (0.06$\pm$0.02) \\
\hline
\multirow{2}{*}{S$_{\rm E}$} & Disturbed &87 (113) & 0.05$\pm$0.04 (0.05$\pm$0.03) &0.25$\pm$0.08 (0.27$\pm$0.07) &0.70$\pm$0.08 (0.68$\pm$0.07) \\
                        & Undisturbed &97 (145) & 0.22$\pm$0.07 (0.27$\pm$0.06) &0.38$\pm$0.08 (0.39$\pm$0.07) &0.40$\pm$0.08 (0.34$\pm$0.06) \\
\hline
\multirow{2}{*}{S$_{\rm L}$} & Disturbed &47 (57) & 0.00$\pm$0.00 (0.00$\pm$0.00) &0.02$\pm$0.03 (0.03$\pm$0.03) &0.98$\pm$0.03 (0.97$\pm$0.03) \\
                        & Undisturbed &67 (89) & 0.07$\pm$0.05 (0.12$\pm$0.06) &0.13$\pm$0.07 (0.15$\pm$0.06) &0.79$\pm$0.08 (0.73$\pm$0.08) \\
\hline \hline
\multirow{2}{*}{BT1} & Disturbed &54 (73) & 0.26$\pm$0.10 (0.28$\pm$0.09) &0.43$\pm$0.11 (0.44$\pm$0.10) &0.31$\pm$0.10 (0.28$\pm$0.09) \\
                        & Undisturbed &261 (463) & 0.53$\pm$0.05 (0.58$\pm$0.04) &0.41$\pm$0.05 (0.38$\pm$0.04) &0.06$\pm$0.02 (0.05$\pm$0.02) \\
\hline
\multirow{2}{*}{BT2} & Disturbed &65 (88) & 0.09$\pm$0.06 (0.10$\pm$0.05) &0.52$\pm$0.10 (0.54$\pm$0.09) &0.38$\pm$0.10 (0.37$\pm$0.08) \\
                        & Undisturbed &207 (335) & 0.47$\pm$0.06 (0.50$\pm$0.04) &0.45$\pm$0.06 (0.43$\pm$0.04) &0.08$\pm$0.03 (0.07$\pm$0.02) \\
\hline
\multirow{2}{*}{BT3} & Disturbed &63 (82) & 0.14$\pm$0.07 (0.15$\pm$0.06) &0.35$\pm$0.10 (0.37$\pm$0.09) &0.51$\pm$0.10 (0.48$\pm$0.09) \\
                        & Undisturbed &103 (165) & 0.39$\pm$0.08 (0.44$\pm$0.06) &0.39$\pm$0.08 (0.38$\pm$0.06) &0.22$\pm$0.07 (0.18$\pm$0.05) \\
\hline
\multirow{2}{*}{BT3} & Disturbed &71 (88) & 0.01$\pm$0.02 (0.02$\pm$0.03) &0.07$\pm$0.05 (0.08$\pm$0.05) &0.92$\pm$0.05 (0.90$\pm$0.05) \\
                        & Undisturbed &82 (110) & 0.09$\pm$0.05 (0.11$\pm$0.05) &0.18$\pm$0.07 (0.22$\pm$0.06) &0.73$\pm$0.08 (0.67$\pm$0.07) \\
\hline \hline

\multicolumn{6}{l}{$^a$ B/T group: BT1 (B/T $>$0.7); BT2 (0.4$<$B/T$<$0.7); BT3 (0.2$<$B/T$<$0.4); BT4 (B/T$<$0.2).}\\
\multicolumn{6}{l}{$^b$ The number of galaxies for each group. Parentheses are the corrected ones for the NUV completeness (see Section 2).}\\
\multicolumn{6}{l}{$^c$ The fraction of optical colour group. Parentheses are the corrected ones for the NUV completeness (see Section 2).}\\

        \end{tabular}
        \label{tab:optcol} 
      \end{table*}

\section{The UV-optical colour-magnitude relation}
\label{sec:nuv}
\begin{figure}
       \centering
       \includegraphics[width=0.52\textwidth]{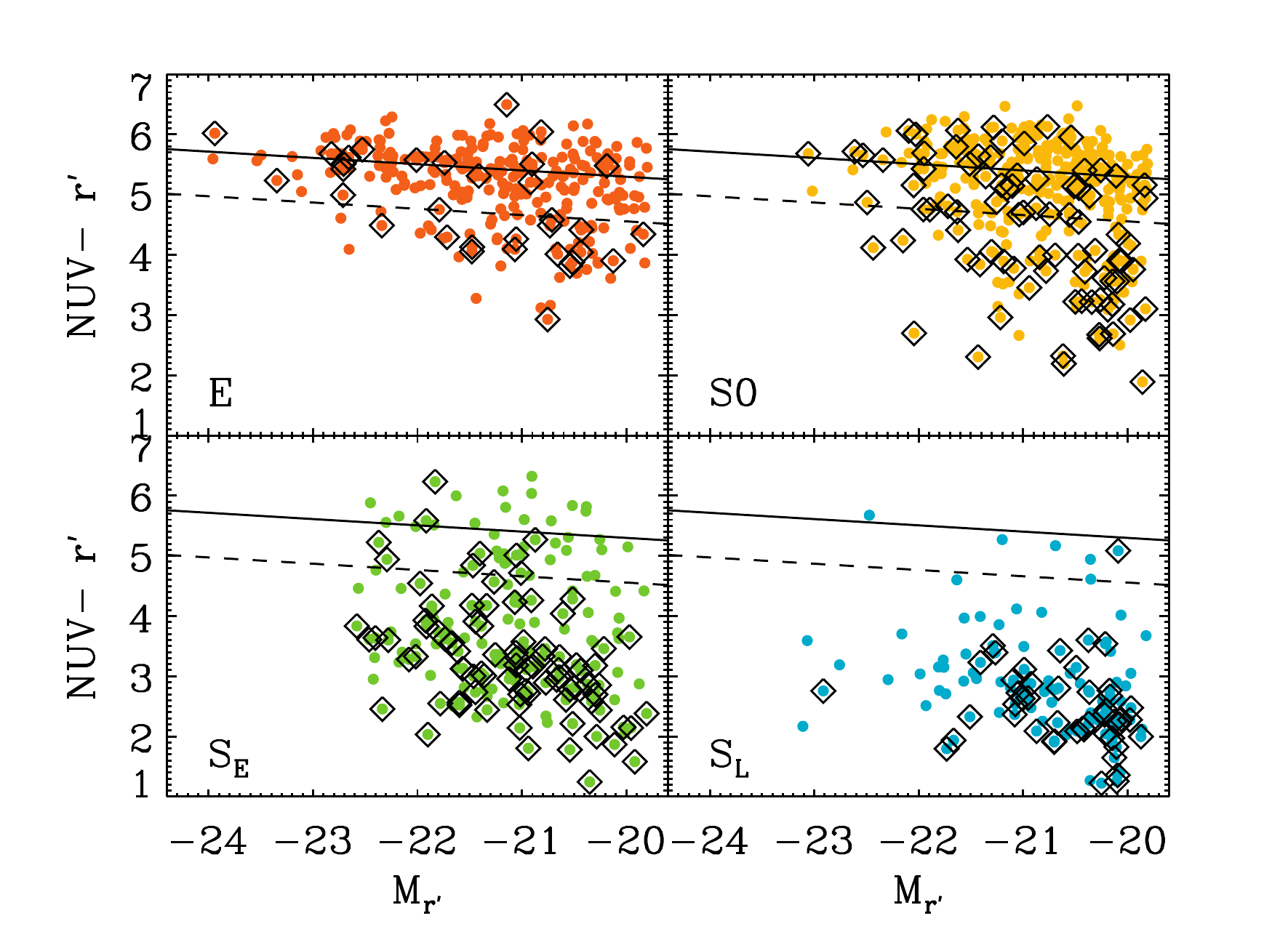}
       \caption[The composite NUV$-\rmag$ CMD for each morphology.]
      {The composite NUV$-\rmag$ CMD for each morphology. Solid and dashed lines indicate the RS and the sequence $2\sigma$ below the RS based on elliptical galaxies (E), respectively. Morphologically-disturbed galaxies which are denoted by diamonds show bluer colour compared to undisturbed galaxies in the same morphology.}
     \label{cmdnuv}
     \end{figure}

\begin{figure}
       \centering
       \includegraphics[width=0.52\textwidth]{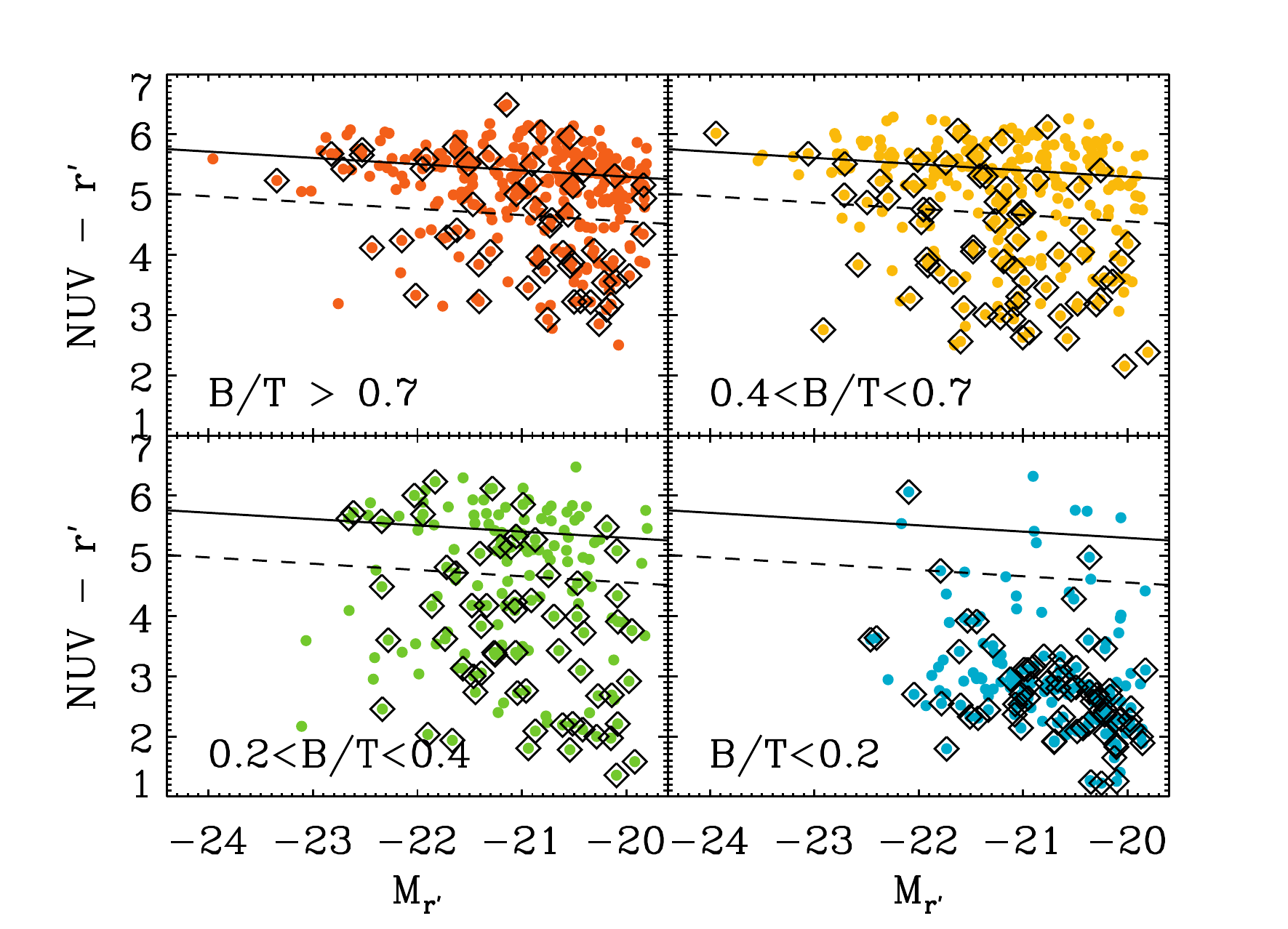}
       \caption[The composite NUV$-\rmag$ CMD for each B/T]
      {The composite NUV$-\rmag$ CMD for each B/T subgroup. Details are same as Figure~\ref{cmdnuv}.}
     \label{btnuv}
     \end{figure}

\begin{figure}
       \centering
       \includegraphics[width=0.5\textwidth]{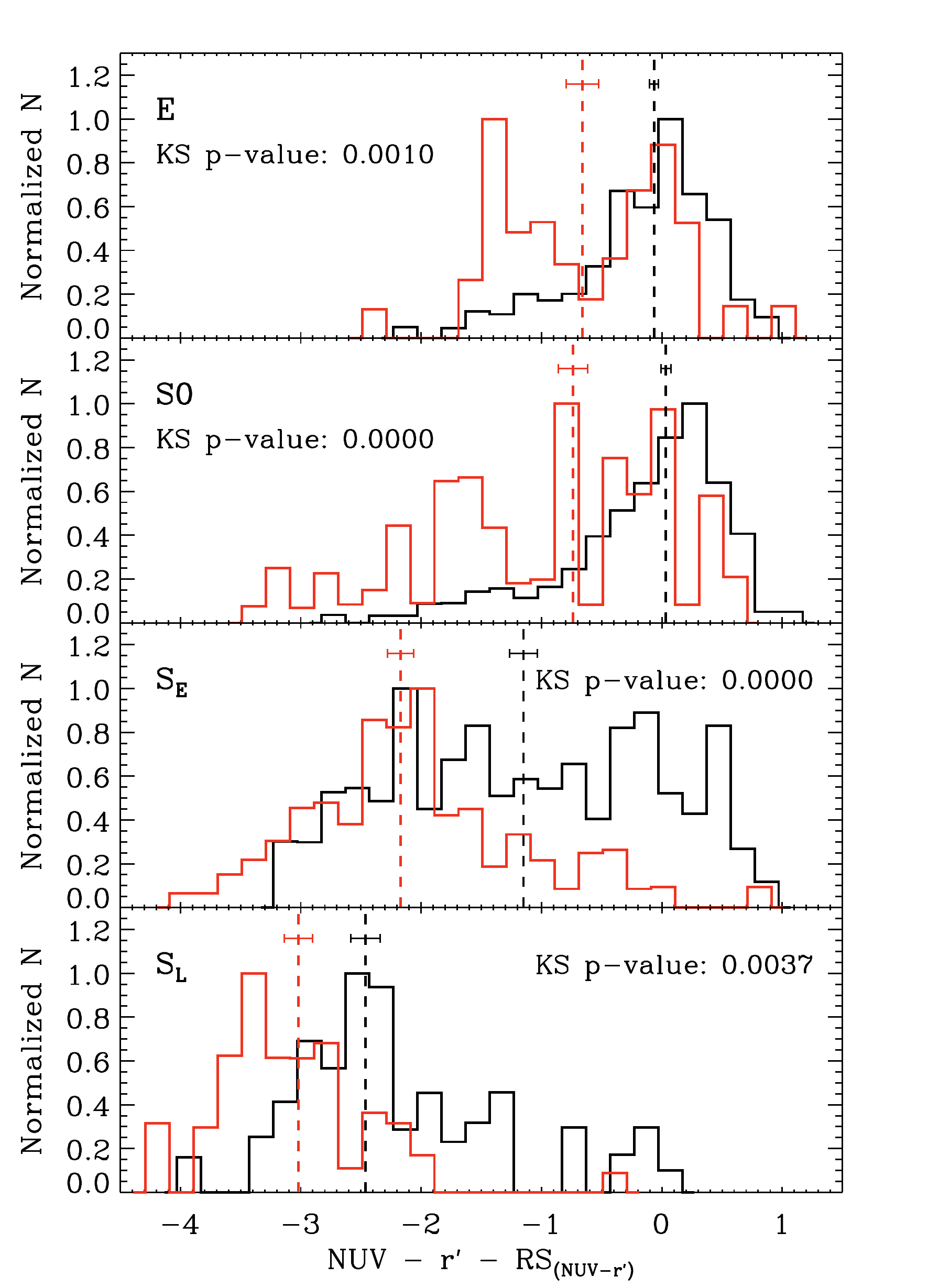}
       \caption[The NUV$-\rmag$ colour distribution]
      {The NUV$-\rmag$ colour distribution corrected for the NUV completeness. Dashed lines indicate median value of colour, suggesting bluer colour distribution of disturbed (red) galaxies compared undisturbed ones (black). Error bars represent standard error of the median. The median p-value from the thousand iterations of the KS test considering NUV$-\rmag$ errors is shown in each panel.}
     \label{cmddis2}
     \end{figure}

 As expected from the optical CMD, morphologically-disturbed galaxies seem to have bluer NUV$-\rmag$ colour than undisturbed galaxies (Figures~\ref{cmdnuv} and \ref{btnuv}). We applied a similar strategy to the UV-optical CMDs. The CMD of elliptical galaxies reveals the distinct red-sequence (RS$_{\rm NUV-\rmag}$) of NUV$-\rmag=-0.104\, M_{\rmag}\,+\,3.214$. Table~\ref{tab:nuvcol} shows the fractions of UV-optical colour groups for disturbed and undisturbed galaxies: NUV redder RS galaxies (NRRS; galaxies redder than the RS$_{\rm NUV-\rmag}$), NUV bluer RS galaxies (NBRS; galaxies bluer than the RS$_{\rm NUV-\rmag}$ and redder than the sequence $2\sigma$ below from the RS$_{\rm NUV-\rmag}$), and NUV blue galaxies (NBG; galaxies bluer than the sequence $2\sigma$ below from the RS$_{\rm NUV-\rmag}$). We found that 24$\pm3$\% and 83$\pm4$\% of early- and late-type galaxies are categorised as the NBG group, respectively. In addition, the NUV$-\rmag$ distribution of disturbed galaxies is weighted toward blue NUV$-\rmag$ colour compared to undisturbed ones in all morphologies (Figure~\ref{cmddis2}). 

The blue NUV$-\rmag$ colour of post-merger galaxies was reported in Sheen et al. (2016). They found that 36\% of post-merger and 26\% of normal galaxies present NUV$-\rmag<5$ based on 263 early-type galaxies in the optical RS. For comparison, we examined statistics based only on 552 early-type galaxies (E and S0) in the optical RS (RRS and BRS), and found that 37$\pm9$\% (32/87) of disturbed and 16$\pm3$\% (74/465) of undisturbed early types in RS show NUV$-\rmag<5$. We found comparable results following our NBG criterion that 36$\pm8$\% (31/87) of disturbed and 15$\pm3$\% (68/465) of undisturbed early types in RS are classified into the NBG group. The NUV-$\rmag$ blue galaxies are detected more often in merger-related galaxies in both Sheen et al. (2012) and this study. 

Our sample including optically blue galaxies (BG) increases NBG fractions especially in disturbed galaxies. As shown in Table~\ref{tab:optcol}, 27\% of disturbed early-type galaxies lie away from the RS showing blue $\gmag-\rmag$ colour. Including optically blue galaxies (BG), we discovered that 50$\pm8$\% (60/119) of disturbed and 17$\pm3$\% (85/489) of undisturbed early-type galaxies are classified into NBG. Late-type galaxies also show considerable differences in colour between disturbed and undisturbed populations (Table~\ref{tab:nuvcol} and Figure~\ref{cmddis2}). The statistics based on B/T groups yield a comparable result to the one based on morphology groups (Table~\ref{tab:nuvcol}). These results give us a hint that galaxies which experienced recent mergers have younger stellar populations compared to their counterparts.

          \begin{table*}
        \centering
        \caption[The fractions of UV colour groups]{The fractions of UV colour groups}
        \begin{tabular}{lllllll}
        \hline \hline
      Morphology$^{a}$ &   Merger    & N$^b$ & NRRS$^c$ & NBRS$^c$ & NBG$^c$\\
        \hline \hline
\multirow{2}{*}{E} & Disturbed &34 (47) & 0.29$\pm$0.13 (0.30$\pm$0.11) &0.24$\pm$0.12 (0.24$\pm$0.10) &0.47$\pm$0.14 (0.46$\pm$0.12) \\
                       & Undisturbed &237 (395) & 0.43$\pm$0.05 (0.44$\pm$0.04) &0.40$\pm$0.05 (0.39$\pm$0.04) &0.17$\pm$0.04 (0.17$\pm$0.03) \\
\hline
\multirow{2}{*}{S0} & Disturbed &85 (115) & 0.24$\pm$0.08 (0.25$\pm$0.07) &0.25$\pm$0.08 (0.25$\pm$0.07) &0.52$\pm$0.09 (0.50$\pm$0.08) \\
                       & Undisturbed &252 (445) & 0.51$\pm$0.05 (0.53$\pm$0.04) &0.31$\pm$0.05 (0.31$\pm$0.04) &0.18$\pm$0.04 (0.16$\pm$0.03) \\
\hline
\multirow{2}{*}{S$_{\rm E}$} & Disturbed &87 (113) & 0.02$\pm$0.03 (0.03$\pm$0.02) &0.08$\pm$0.05 (0.08$\pm$0.04) &0.90$\pm$0.05 (0.89$\pm$0.05) \\
                       & Undisturbed &97 (145) & 0.15$\pm$0.06 (0.18$\pm$0.05) &0.21$\pm$0.07 (0.24$\pm$0.06) &0.64$\pm$0.08 (0.59$\pm$0.07) \\
\hline
\multirow{2}{*}{S$_{\rm L}$} & Disturbed &47 (57) & 0.00$\pm$0.00 (0.00$\pm$0.00) &0.02$\pm$0.03 (0.03$\pm$0.03) &0.98$\pm$0.03 (0.97$\pm$0.03) \\
                       & Undisturbed &67 (89) & 0.01$\pm$0.02 (0.01$\pm$0.02) &0.06$\pm$0.05 (0.10$\pm$0.05) &0.93$\pm$0.05 (0.89$\pm$0.05) \\
\hline \hline
\multirow{2}{*}{BT1} & Disturbed &54 (73) & 0.24$\pm$0.10 (0.26$\pm$0.08) &0.22$\pm$0.09 (0.24$\pm$0.08) &0.54$\pm$0.11 (0.51$\pm$0.10) \\
                      & Undisturbed &261 (463) & 0.44$\pm$0.05 (0.46$\pm$0.04) &0.38$\pm$0.05 (0.38$\pm$0.04) &0.18$\pm$0.04 (0.17$\pm$0.03) \\
\hline
\multirow{2}{*}{BT2} & Disturbed &65 (88) & 0.14$\pm$0.07 (0.14$\pm$0.06) &0.23$\pm$0.09 (0.23$\pm$0.07) &0.63$\pm$0.10 (0.63$\pm$0.08) \\
                       & Undisturbed &207 (335) & 0.44$\pm$0.06 (0.46$\pm$0.04) &0.33$\pm$0.05 (0.33$\pm$0.04) &0.23$\pm$0.05 (0.21$\pm$0.04) \\
\hline
\multirow{2}{*}{BT3} & Disturbed &63 (82) & 0.14$\pm$0.07 (0.16$\pm$0.07) &0.13$\pm$0.07 (0.13$\pm$0.06) &0.73$\pm$0.09 (0.71$\pm$0.08) \\
                    & Undisturbed &103 (165) & 0.35$\pm$0.08 (0.37$\pm$0.06) &0.26$\pm$0.07 (0.29$\pm$0.06) &0.39$\pm$0.08 (0.34$\pm$0.06) \\
\hline
\multirow{2}{*}{BT4} & Disturbed &71 (88) & 0.01$\pm$0.02 (0.02$\pm$0.02) &0.03$\pm$0.03 (0.03$\pm$0.03) &0.96$\pm$0.04 (0.95$\pm$0.04) \\
                        & Undisturbed &82 (110) & 0.07$\pm$0.05 (0.09$\pm$0.05) &0.04$\pm$0.03 (0.05$\pm$0.03) &0.89$\pm$0.06 (0.86$\pm$0.05) \\
\hline \hline

\multicolumn{6}{l}{$^a$ B/T group: BT1 (B/T $>$0.7); BT2 (0.4$<$B/T$<$0.7); BT3 (0.2$<$B/T$<$0.4); BT4 (B/T$<$0.2).}\\
\multicolumn{6}{l}{$^b$ The number of galaxies for each group. Parentheses are the corrected ones for the NUV completeness (see Section 2).}\\
\multicolumn{6}{l}{$^c$ The fraction of NUV colour group. Parentheses are the corrected ones for the NUV completeness (see Section 2).}\\

        \end{tabular}

        \label{tab:nuvcol} 
      \end{table*}

\begin{table}
\centering
\caption[NUV colour offsets]{The $NUV-\rmag$ colour offsets}
\begin{tabular}{llll c c c} \hline \hline
Cluster & N$_{\rm U}^{a}$ & N$_{\rm D}^{b}$ & $\Delta \overline{(NUV-\rmag)}^{c}$  \\
\hline
Abell 116 & 29 (47)      & 9 (11)    & 1.21$\pm$0.28 (1.26$\pm$0.30)\\
Abell 646 & 30 (47)      & 11 (14)  & 0.82$\pm$0.19 (0.84$\pm$0.20)\\
Abell 655 & 68 (102)    & 36 (45)  & 0.88$\pm$0.04 (0.97$\pm$0.06)\\
Abell 667 & 32 (49)      & 16 (20)  & 1.29$\pm$0.10 (1.25$\pm$0.11)\\
Abell 690 & 44 (76)      & 22 (27)  & 1.30$\pm$0.13 (1.36$\pm$0.14)\\
Abell 1126 & 32 (57)     & 18 (22) & 1.38$\pm$0.10 (1.45$\pm$0.13)\\
Abell 1139 & 35 (58)     & 11 (15) & 1.26$\pm$0.11 (1.22$\pm$0.15)\\
Abell 1146 & 53 (89)     & 25 (32) & 0.73$\pm$0.10 (0.73$\pm$0.10)\\
Abell 1278 & 18 (26)     & 9 (13)   & 1.24$\pm$0.19 (1.12$\pm$0.18)\\
Abell 2061 & 128 (209) & 43 (55) & 1.22$\pm$0.10 (1.25$\pm$0.10)\\
Abell 2249 & 120 (202) & 36 (53) & 0.56$\pm$0.11 (0.60$\pm$0.10)\\
Abell 2589 & 52 (95)     & 9 (13)   & 1.23$\pm$0.41 (1.11$\pm$0.34)\\
Abell 3574 & 8 (11)       & 4 (6)     & -0.19$\pm$0.40 (-0.04$\pm$0.32)\\
Abell 3659 & 4 (6)         & 4 (5)     & 1.36$\pm$0.34 (1.32$\pm$0.24)\\
\hline
Total          & 653 (1073) & 253 (331) & 1.09$\pm$0.04 (1.11$\pm$0.02)\\
\hline
\hline
\multicolumn{4}{l}{$^a$ Number of undisturbed galaxies. Parentheses are the} \\
\multicolumn{4}{l}{corrected ones for the NUV completeness (see Section 2).}\\
\multicolumn{4}{l}{$^b$ Number of disturbed galaxies. Parentheses are the corrected} \\
\multicolumn{4}{l}{ones for the NUV completeness (see Section 2).}\\         
\multicolumn{4}{l}{$^c$ Difference in mean ${ (NUV-\rmag) }$ colour of disturbed galaxies}\\
\multicolumn{4}{l}{to undisturbed one.}\\
\end{tabular}
\label{tab:off}
\end{table} 

Table~\ref{tab:off} shows the difference in mean NUV$-\rmag$ colour between disturbed and undisturbed galaxies in each cluster. Disturbed galaxies are bluer than undisturbed galaxies by 0.56 - 1.38 in mean NUV$-\rmag$ colour except for Abell 3574 which has the lowest cluster mass ($1.9 \times 10^{14} M_{\odot}$) from our sample and has only two red sequence galaxies. Although there is a variation in colour offsets -- probably because of small number statistics and the difference in population of each cluster (i.e. early-type fractions) -- we found bluer disturbed galaxies in most of our sample clusters with various cluster masses and populations.

\section{Phase-space analysis}
 The optical and UV colour distributions clearly show that merger-related galaxies have more blue (i.e. young) populations than undisturbed galaxies. Before discussing the impact of mergers, we need to examine the environmental dependence of the galaxy populations. Galaxy clusters are the sites where the quenching of star formation actively occurs by their potential (ram-pressure stripping, e.g., Gunn \& Gott 1972; Smith et al. 2016). Numerical simulations predict that the time that a galaxy spent in a cluster is a critical factor determining the amount of quenching of star formation (Jung et al. 2018). Observations have also supported this by presenting the increasing numbers of star-forming populations with clustercentric distance (Haines et al. 2015). In Oh et al. (2018), we conclude that galaxies with merger-related features (OM and PM subgroups in the paper) are found more on the cluster outskirts and, therefore, recently fell into the cluster environment. In these regards, the younger populations found in disturbed galaxies may indicate that they have been less influenced by the quenching process than undisturbed galaxies.

\begin{figure}
       \centering
       \includegraphics[width=0.5\textwidth]{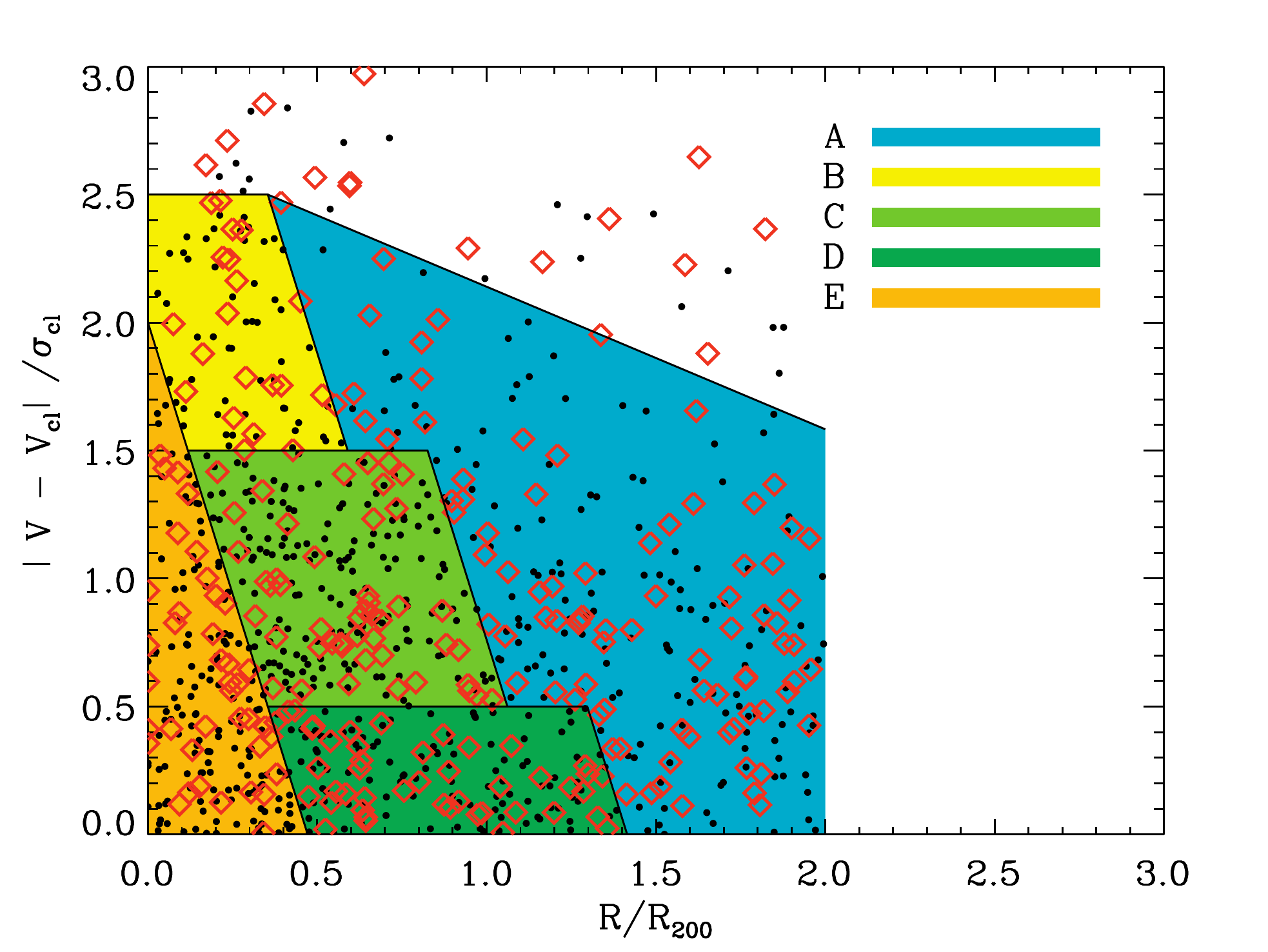}
       \caption[The mass-colour diagram]
      {The phase-space diagram. The grids separating five regions (A-E) are adopted from Rhee et al. (2017), but modified to be the one based on $R/R_{200}$. Both disturbed (red diamonds) and undisturbed (black circles) galaxies are spread over the regions.}
     \label{phase}
     \end{figure}

Recently, Rhee et al. (2017) studied the location of galaxies on the phase-space plane according to the time since infall into a cluster based on the cosmological hydrodynamic N-body simulations by Choi \& Yi (2017). They divided cluster galaxies into five subgroups according to the time since infall in to a cluster ($t_{\rm inf}$): Interlopers, First Infallers (not fallen yet), Recent Infallers ($0 <t_{\rm inf}< 3.63$ Gyrs), Intermediate Infallers ($3.63 <t_{\rm inf}< 6.45$ Gyrs), and Ancient Infallers ($6.45 <t_{\rm inf}< 13.7$ Gyrs). Then, they analysed the number density of each subgroup in five regions on the phase-space diagram. Figure~\ref{phase} is a projected phase-space diagram for our sample galaxies with five distinct regions based on $t_{\rm inf}$, as illustrated in Figure 6 of Rhee et al. (2017). Five regions (A-E) in Figure~\ref{phase} have been modified considering the difference between R$_{\rm vir}$ (Rhee et al. 2017) and R$_{200}$ (this study).

    \begin{table*}
        \centering
        \caption[Information of Phase-Space regions]{Information of Phase-Space regions}
        \begin{tabular}{cccccccc}
        \hline \hline
        Region &  $\overline{t_{\rm inf}}^{a}$ & $f_{FR}^{b}$ & $N_{total}^{c}$& $P_{disburbed}^{d}$ [\%] &$P_{E+S0}^{e}$ [\%]&  $\overline{(NUV-\rmag-RS)_{(E+S0)}}^{f}$&$\overline{(NUV-\rmag-RS)_{(S_E+S_L)}}^{g}$\\
        \hline \hline 
A &     0.44  &0.86& 232 (332) & 36$\pm$5 (33$\pm$4)   & 50$\pm$5 (54$\pm$4)  & -0.66$\pm$0.09 (-0.60$\pm$0.06) & -2.18$\pm$0.09 (-2.10$\pm$0.08)\\
\hline
B &    3.98 &0.46 & 74 (118) & 27$\pm$8 (22$\pm$6) & 73$\pm8$ (78$\pm6$) & -0.23$\pm0.10$ (-0.18$\pm$0.07) & -2.50$\pm0.27$ (-2.35$\pm$0.25) \\
\hline
C &     3.49&0.51&  214 (345) & 22$\pm5$ (18$\pm3$) & 70$\pm5$ (73$\pm4$) & -0.21$\pm0.06$ (-0.16$\pm$0.04) & -1.63$\pm0.14$ (-1.45$\pm$0.12) \\
\hline 
D &    3.09 &0.47& 158 (236) & 28$\pm6$ (24$\pm5$) & 63$\pm6$ (68$\pm5$) & -0.35$\pm0.08$ (-0.29$\pm$0.06) & -2.18$\pm0.12$ (-2.08$\pm$0.10) \\
\hline
E  &    5.82 &0.21& 185 (310) & 21$\pm5$ (17$\pm4$) &  86$\pm4$ (88$\pm3$) & -0.13$\pm0.05$ (-0.10$\pm$0.01) & -1.15$\pm0.28$ (-0.95$\pm$0.15)\\

\hline \hline

\multicolumn{8}{l}{$^a$ Mean time in Gyrs since infall into a cluster for each phase-space region (Rhee et al. 2017).}\\
\multicolumn{8}{l}{$^b$ Fraction of First and Recent Infallers ($t_{\rm inf}<3.63$ Gyrs) for each phase-space region (Rhee et al. 2017).}\\          
\multicolumn{8}{l}{$^c$ Number of galaxies within the phase-space region. Parentheses are the corrected ones for the NUV completeness (see Section 2).}\\
\multicolumn{8}{l}{$^d$ Percentage of disturbed galaxies in the phase-space region.}\\
\multicolumn{8}{l}{$^e$ Percentage of early-type (E+S0) galaxies in the phase-space region.}\\
\multicolumn{8}{l}{$^f$ Mean NUV$-\rmag$ colour of early-type (E+S0) galaxies corrected for the RS$_{(NUV-\rmag)}$ where ${(NUV-\rmag-RS_{NUV-\rmag})}$=0.}\\
\multicolumn{8}{l}{$^g$ Mean NUV$-\rmag$ colour of late-type (S$\rm_E$+S$\rm_L$) galaxies corrected for the RS$_{(NUV-\rmag)}$.}\\

 \end{tabular}

        \label{tab:phase} 
      \end{table*}

      \begin{figure*}
       \centering
       \includegraphics[width=1.\textwidth]{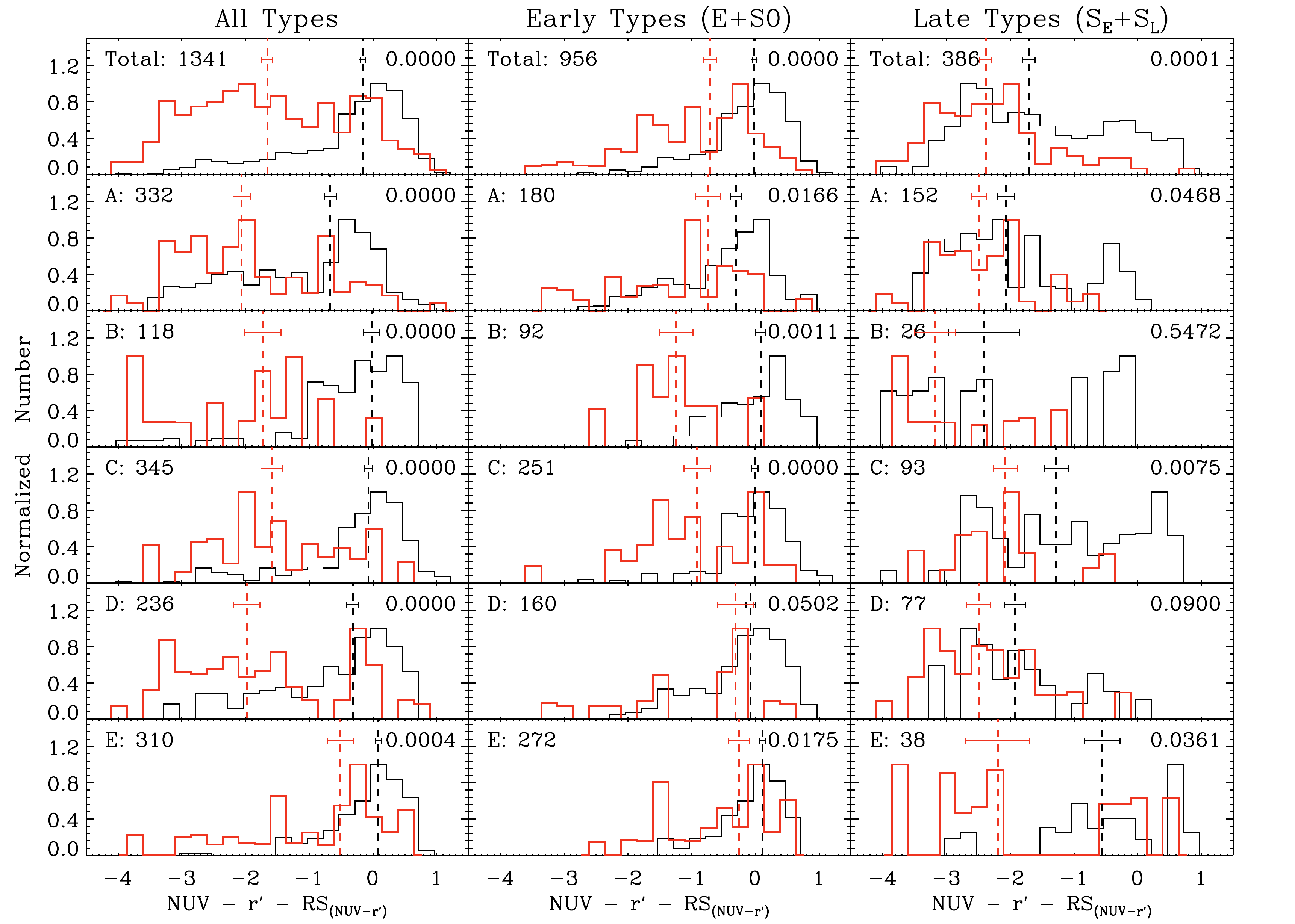}
       \caption[The NUV$-\rmag$ colour distributions]
      {The NUV$-\rmag$ colour distributions of disturbed (red) and undisturbed (black) galaxies corrected for the NUV completeness. Regions A-E are adopted from Figure~\ref{phase}. Dashed lines indicate the median colour of each group. Error bars represent standard error of the median. The median p-value from the thousand iterations of the KS test considering NUV$-\rmag$ errors is shown in the top right. Disturbed galaxies show a bluer colour distribution with undisturbed ones in all regions and morphologies.}
     \label{phasehist}
     \end{figure*}

      \begin{figure}
       \centering
       \includegraphics[width=0.5\textwidth]{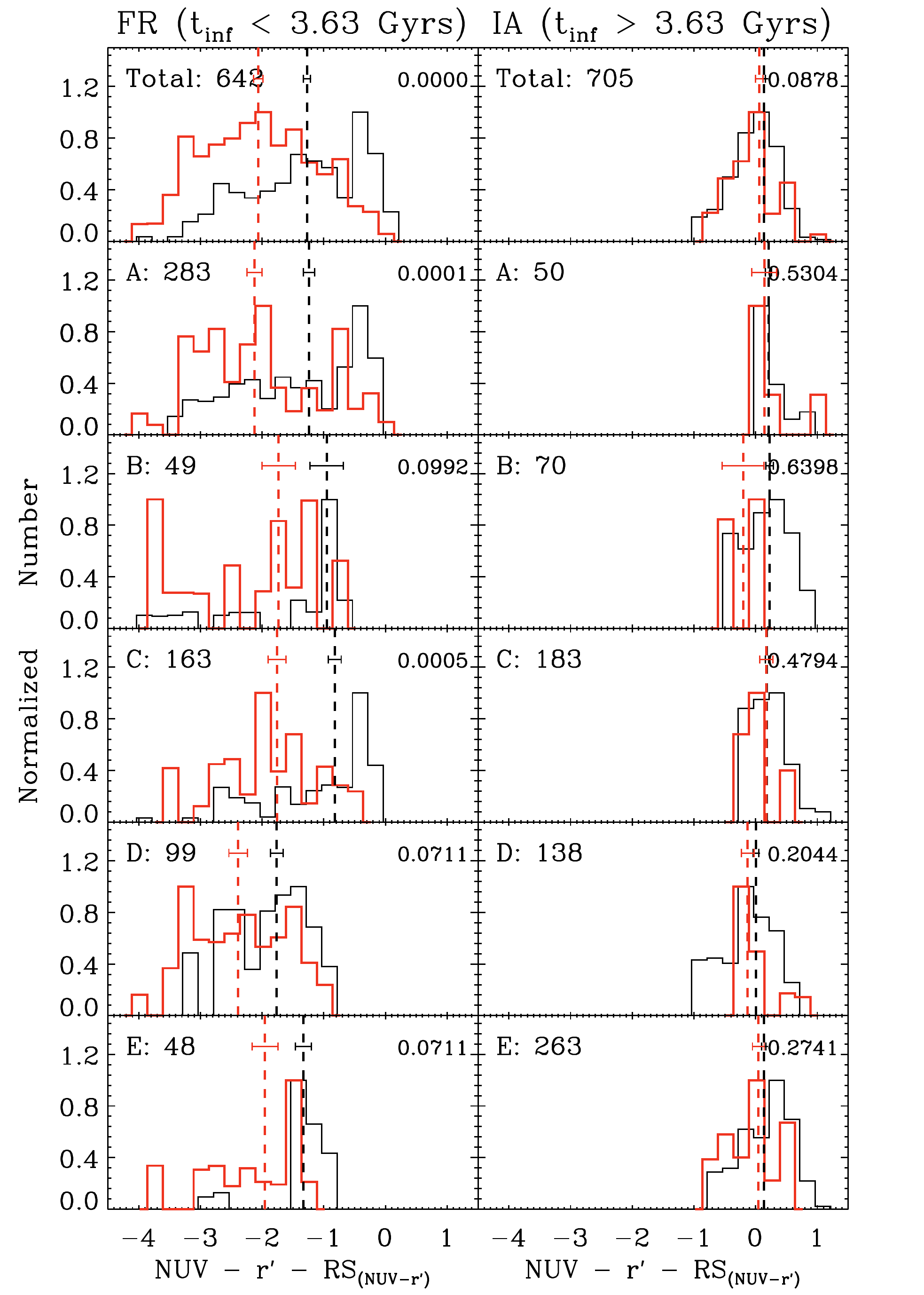}
       \caption[The NUV$-\rmag$ colour distributions]
      {Same as Figure~\ref{phasehist} but galaxies have been divided based on the time since infall ($t_{\rm inf}$) with an extreme assumption that FR and IA samples are the bluest and reddest galaxies in each region. The FR sample shows a difference in colour between disturbed undisturbed galaxies.}
     \label{tphasehist}
     \end{figure}

In Table~\ref{tab:phase}, we present the fractions of disturbed and early-type (E+S0) galaxies, and the mean NUV$-\rmag$ colour for early and late types in each region. Populations of regions A (cluster outskirts) and E (core) show a clear difference in $f_{disturbed}$, $f_{E+S0}$, and $\overline{\rm NUV-\rmag-RS_{\rm NUV-\rmag}}$: Region E is dominated by NUV-red, early-type, and/or undisturbed galaxies, whereas Region A displays a significant fraction of NUV-blue, late-type and/or disturbed populations. Regions B, C, and D show intermediate properties between A and E, and we could not find a significant difference between the regions, possibly due to a projection effect on the line-of-sight phase-space analysis.

Significantly bluer colour distributions are detected in disturbed galaxies in all geometrically-selected regions (first column of Figure~\ref{phasehist}). Although galaxies in Region E (cluster core) are mostly quenched populations near the RS, we found blue tails in disturbed galaxies. Late-type galaxies are naturally expected to have high levels of star formation and gas contents, therefore, the bluer colour distribution of disturbed galaxies could be the result of having more late-type galaxies than undisturbed ones. Indeed, we found a much higher fraction of late-type galaxies in disturbed galaxies (52\%) than in undisturbed galaxies (25\%).

We found differences in colour distributions between disturbed and undisturbed galaxies, even for fixed galaxy types (second and third columns of Figure~\ref{phasehist}). In region A, the median colour of disturbed galaxies is bluer by 0.4 than that of undisturbed galaxies, which indicates an enhancement in specific star formation rate of roughly by 85\% (Salim et al. 2009). Following the results of Rhee et al. (2017), 86\% of the galaxies in Region A are first (77\%) and recent (9\%) infallers, which are less likely to be affected by environmental quenching. Therefore, we suspect that the primary drivers making disturbed galaxies bluer in the cluster outskirts are mergers. 

Galaxies in Regions B, C, and D have a range of $t_{\rm inf}$, which can introduce the colour difference between disturbed and undisturbed galaxies. However, our analysis for the fixed morphology as well as geometrical regions may partially offset the effect of $t_{\rm inf}$; the range of $t_{\rm inf}$ might be narrow for the same morphological type in the same region. Additionally, disturbed early-type galaxies show bluer mean NUV$-\rmag$ colour roughly by 1 compared to undisturbed galaxies in Regions B and C. This difference is two times larger than the difference in Regions A ($\overline{(NUV-\rmag-RS_{NUV-\rmag})_{(E+S0)}}$=-0.66) and E ($\overline{(NUV-\rmag-RS_{NUV-\rmag})_{(E+S0)}}$=-0.13), which suggests additional mechanisms (i.e. mergers) making disturbed early-type galaxies bluer. In Region E (cluster core), we mostly found quenched populations near the RS, but we also found measurable differences in mean NUV$-\rmag$ colours between disturbed and undisturbed galaxies.

In Table~\ref{tab:phase}, we also present the prediction for the fraction of first and recent infallers ($f_{FR}$) in each region adopted from Rhee et al. (2017). The fraction of intermediate and ancient infallers ($f_{IA}$) is ($1-f_{FR}$) ignoring the fraction of interlopers. We categorised the sample galaxies based on $t_{\rm inf}$ with an extreme assumption that FR ($t_{\rm inf}<3.63$ Gyrs) and IA  ($t_{\rm inf}>3.63$ Gyrs) samples are the bluest and reddest galaxies in each region. In Figure~\ref{tphasehist}, the FR sample clearly shows the difference in NUV$-\rmag$ colour between disturbed and undisturbed galaxies, whereas the IA sample shows nearly identical colour distributions between the two groups. Although this is  based on an extreme assumption, we can expect the impact of mergers in galaxy colours at least for the galaxies recently accreted to the cluster environment. 

\section{Mass dependence on the colour offsets}
\begin{figure}
       \centering
       \includegraphics[width=0.5\textwidth]{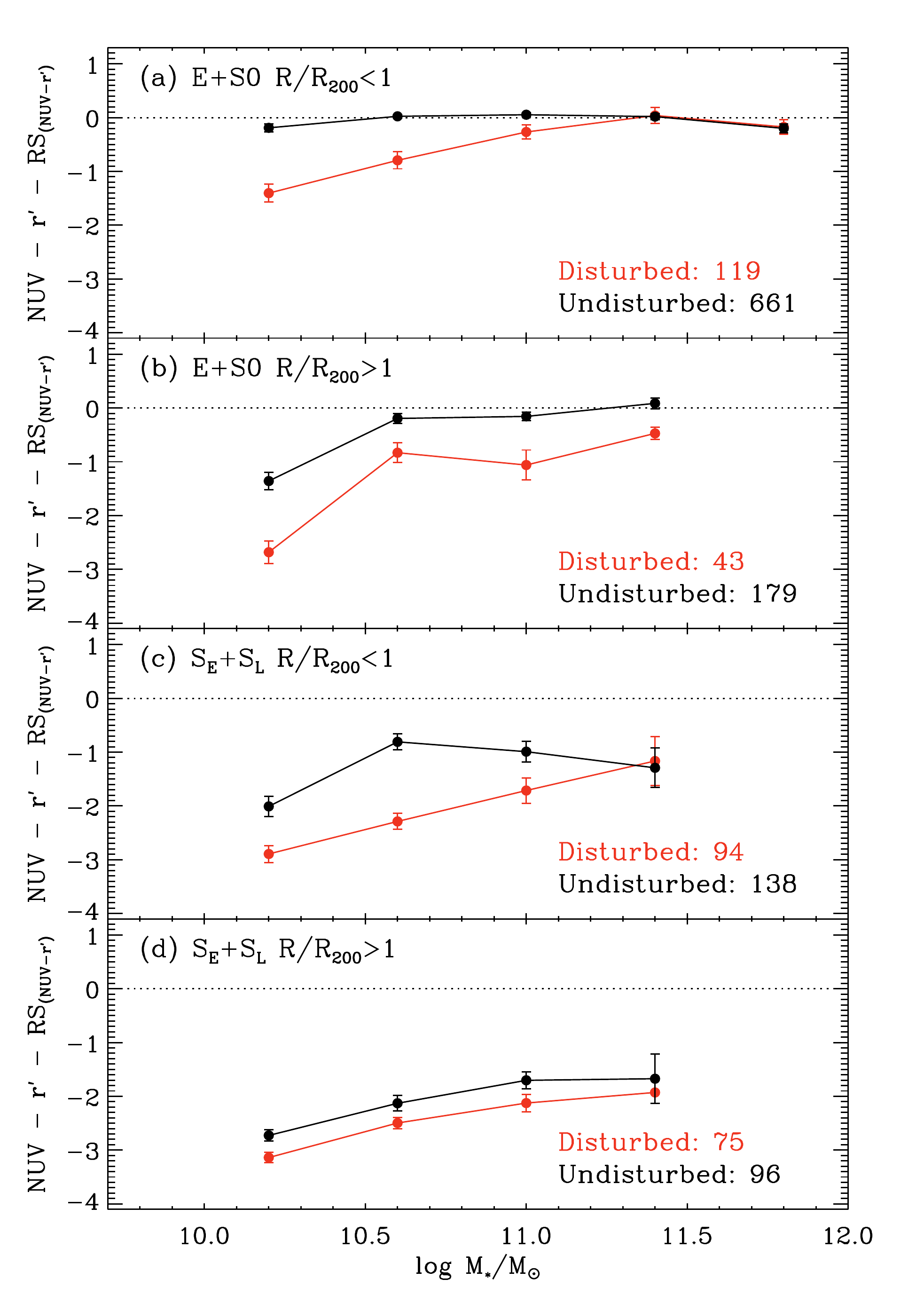}
       \caption[The NUV$-\rmag$ colour distributions]
      {The mean NUV$-\rmag$ colour to stellar mass. Less massive galaxies show more prominent colour differences between disturbed and undisturbed galaxies within $R_{200}$.}
     \label{mass}
     \end{figure}

The stellar mass of galaxies is one other factor to be considered because there is a dependence of star formation on stellar mass (e.g. Kauffmann et al. 2003; Juneau et al. 2005; Zheng, Bell, \& Papovich 2007). We present the mass dependence of mean NUV$-\rmag$ colour in Figure~\ref{mass}. When $R/R_{200} > 1$, there is a clear dependence of mean NUV$-\rmag$ colour on stellar mass in both early- and late-type morphologies. Disturbed galaxies show bluer mean colours than undisturbed galaxies over the range of the stellar mass. The colour offset between disturbed and undisturbed galaxies does not change to the stellar mass (Figure~\ref{mass} (b) and (d)).

When $R/R_{200} < 1$, we found more clear colour offset between disturbed and undisturbed galaxies, especially in less massive galaxies (Figure~\ref{mass} (a) and (c)). Both disturbed and undisturbed galaxies show redder mean colour within $R_{200}$ under the same mass, and the change in mean colours between $R/R_{200} > 1$ and $R/R_{200} < 1$ is much more prominent in undisturbed galaxies. As a result, the colour offset between disturbed and undisturbed galaxies becomes larger in less massive galaxies, and the most massive bin does not show the colour offset.

These results also confirm that star formation caused by merger events makes disturbed galaxies bluer. Within $R_{200}$, both disturbed and undisturbed galaxies show redder NUV$-\rmag$ colour than $R/R_{200} > 1$ and seem to go through environmental quenching. The environmental quenching seems to have more impact on disturbed galaxies, which makes the colour offset more prominent in less massive galaxies.

\section{Disturbed Features}
     \begin{figure*}
       \centering
       \includegraphics[width=0.97\textwidth]{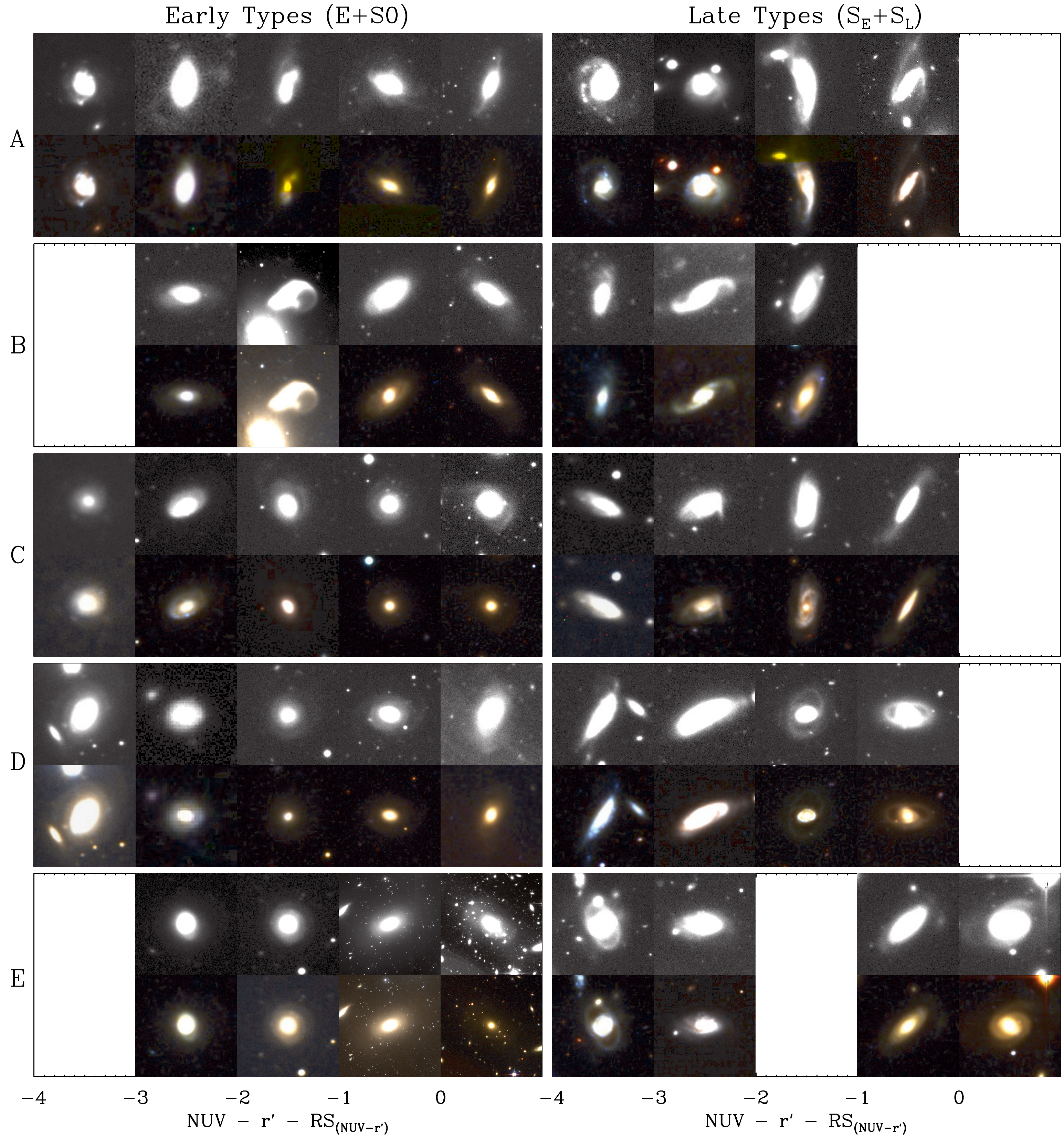}
       \caption[Features of disturbed galaxies]
      {Features of disturbed galaxies in each morphology, NUV$-\rmag$ colour, and phase-space region. The top and bottom row of each region show the $\rmag$-band and (u)gr colour composite images. We found differences in features according to NUV$-\rmag$ colours and phase-space regions.}
     \label{morp}
     \end{figure*}  
     
In Figure~\ref{morp}, we display sample $\rmag$-band and optical colour-composite images of disturbed galaxies for each morphology, NUV$-\rmag-$RS$_{NUV-\rmag}$, and region based on the phase-space diagram. The contrast has been chosen to maximise the identification of features, which may not be the best way to show galaxy morphology. Oh et al. (2018) displays a set of seven cutout images for the visual inspection which allowed us to classify galaxy morphology while considering various contrasts, residuals, and colours.

First, we found a difference in features according to NUV$-\rmag$ colours. When NUV$-\rmag-$RS$_{NUV-\rmag}$ is bluer than -1 (roughly $3\sigma$ below the RS$_{NUV-\rmag}$), we tend to find signatures of star formation such as blue clumps and blue rings which did not appear in the disturbed galaxies near the RS$_{NUV-\rmag}$ (NUV$-\rmag-$RS$_{NUV-\rmag} >$ -1), even for late types. Therefore, we suspect that the disturbed galaxies with blue NUV$-\rmag$ colours have merger-induced star formation through wet mergers. We also barely find features seemingly related to major mergers, especially for blue disturbed galaxies. Features related to major mergers (i.e. ripples and shells) have been found only in nine red and massive early-type galaxies near the cluster centre, which implies that the star formation is mainly caused by minor mergers. This corresponds with previous findings on the importance of minor mergers in the star formation budget (Martin et al. 2018; Kaviraj 2014a, 2014b).

Disturbed galaxies seem to display varying features according to their location in a cluster. Features found in Region A (cluster outskirts) tend to be more violent and irregular in both early- and late-type samples, which suggests that it has not been long since the merger occurred in Region A galaxies. This is consistent with our picture on the origin of disturbed cluster galaxies: the mergers that made disturbed features probably happened before the galaxy accretion into the cluster environment (see also Sheen et al. 2012 and Yi et al. 2013). We could not find a clear tendency between Regions B, C, and D because it is not a time sequence of the galaxy accretion due to orbiting motions of galaxies and the projection effect. In Region E (cluster core), we found a few ripples and shells, and we could not find tidal features which are frequent in Region A. Moreover, only a few galaxies show blue clumps and blue rings in Region E which are the evidence of ongoing star formation.

\section{Summary and Conclusion}
We investigated $\gmag-\rmag$ and NUV-$\rmag$ colours of cluster galaxies with merger-induced morphological disturbance based on the catalogue from the KASI-Yonsei Deep Imaging Survey of Clusters (KYDISC; Oh et al. 2018) targeting 14 clusters at $0.016 < z < 0.145$. This study used 906 galaxies whose NUV magnitude error is lower than 3$\sigma$ detection limit and whose NUV aperture size reasonably traces the one for the $\rmag$ band, out of 1409 spectroscopically-confirmed cluster members brighter than -19.8 in the $r$-band and located within $R < 2R_{200}$.

In the cluster environment, galaxies with the signatures of ongoing or post-merger events show bluer optical and UV colours compared to galaxies without merger signatures in the same morphology or B/T groups. Morphologically-disturbed galaxies not only have more blue populations (Tables~\ref{tab:optcol} and~\ref{tab:nuvcol}), but the entire colour distribution of disturbed galaxies is shifted toward blue optical and UV colours compared to that of undisturbed galaxies (Figures~\ref{cmddis1} and~\ref{cmddis2}). The colour offset between disturbed and undisturbed galaxies has also been detected under the same stellar mass (Figure~\ref{mass}).

The first possible explanation for the bluer colours of disturbed galaxies is the difference in quenching stages between disturbed and undisturbed galaxies. The frequency of disturbed galaxies becomes higher toward the cluster outskirts so, we suspect that disturbed galaxies recently fell into a cluster, and they could hold the visual features to be detected (Oh et al. 2018). The concept of merger relics was discussed by Yi et al. (2013). We conclude that disturbed galaxies have a high chance of spending less time in the cluster environment than undisturbed ones, which also implies disturbed galaxies have less affected by the environmental quenching. 

From the analysis using the phase-space diagram, we detect bluer mean NUV$-\rmag$ colours from the geometrically-selected Region A (cluster outskirts) compared to Region E (cluster core) even in the same morphology (Table~\ref{tab:phase}), which supports the impact of the time since infall into a cluster on galaxy colours. Moreover, we found a more prominent difference in NUV$-\rmag$ colours according to the morphological disturbance from the Regions B, C, and D where galaxies have a range of time since infall into a cluster (Rhee et al. 2017).

However, the difference in the time since infall between two groups cannot fully describe the bluer colours and younger populations of disturbed galaxies. First, we still find bluer colours of disturbed galaxies for the same morphological type (early and late) or the same $t_{\rm inf}$ groups (FR and IA) in all geometrically selected regions (Figures~\ref{phasehist} and \ref{tphasehist}). Second, we found qualitative differences in disturbed features between NUV$-\rmag$ blue and red galaxies: we frequently detect signatures of star formation (e.g. blue clumps) which are seemingly involved with disturbed features (Figure~\ref{morp}).

In summary, cluster galaxies with merger features are bluer (younger) than galaxies that do not show merger signatures for the same morphology, stellar mass, and location. We conclude that the significantly bluer colours of morphologically-disturbed galaxies in the cluster environment are the net result of both the merger-induced changes and the difference in the quenching stage (i.e.  $t_{\rm inf}$). Mergers appear to have triggered star formation as is hinted in Figure~\ref{morp}. Even if star formation has not been activated by mergers, the gas of post-merger galaxies may be concentrated in the centre and thus may be less vulnerable to gas stripping, thus delaying the quenching of their current star forming activities (Smith et al. 2013), which is supported by Figure~\ref{mass} where the environmental quenching seems to have more impact on undisturbed galaxies.

\section*{Acknowledgments}
Parts of this research were conducted by the Australian Research Council Centre of Excellence for All Sky Astrophysics in 3 Dimensions (ASTRO 3D), through project number CE170100013. SKY acknowledges support from the Korean National Research Foundation (2017R1A2A1A05001116). As the PI of the KYDISC project, SKY acted as a corresponding author. MK was supported by the National Research Foundation of Korea (NRF) grant funded by the Korea government (MSIT) (No. 2017R1C1B2002879). YKS acknowledges support from the National Research Foundation of Korea (NRF) grant funded by the Ministry of Science and ICT (No. 2019R1C1C1010279). LCH was supported by the National Key R\&D Program of China (2016YFA0400702) and the National Science Foundation of China (11721303). SO would like to thank DL for the consistent support.

\appendix
\section{conservative sampling}
\begin{figure*}
       \centering
       \includegraphics[width=1.\textwidth]{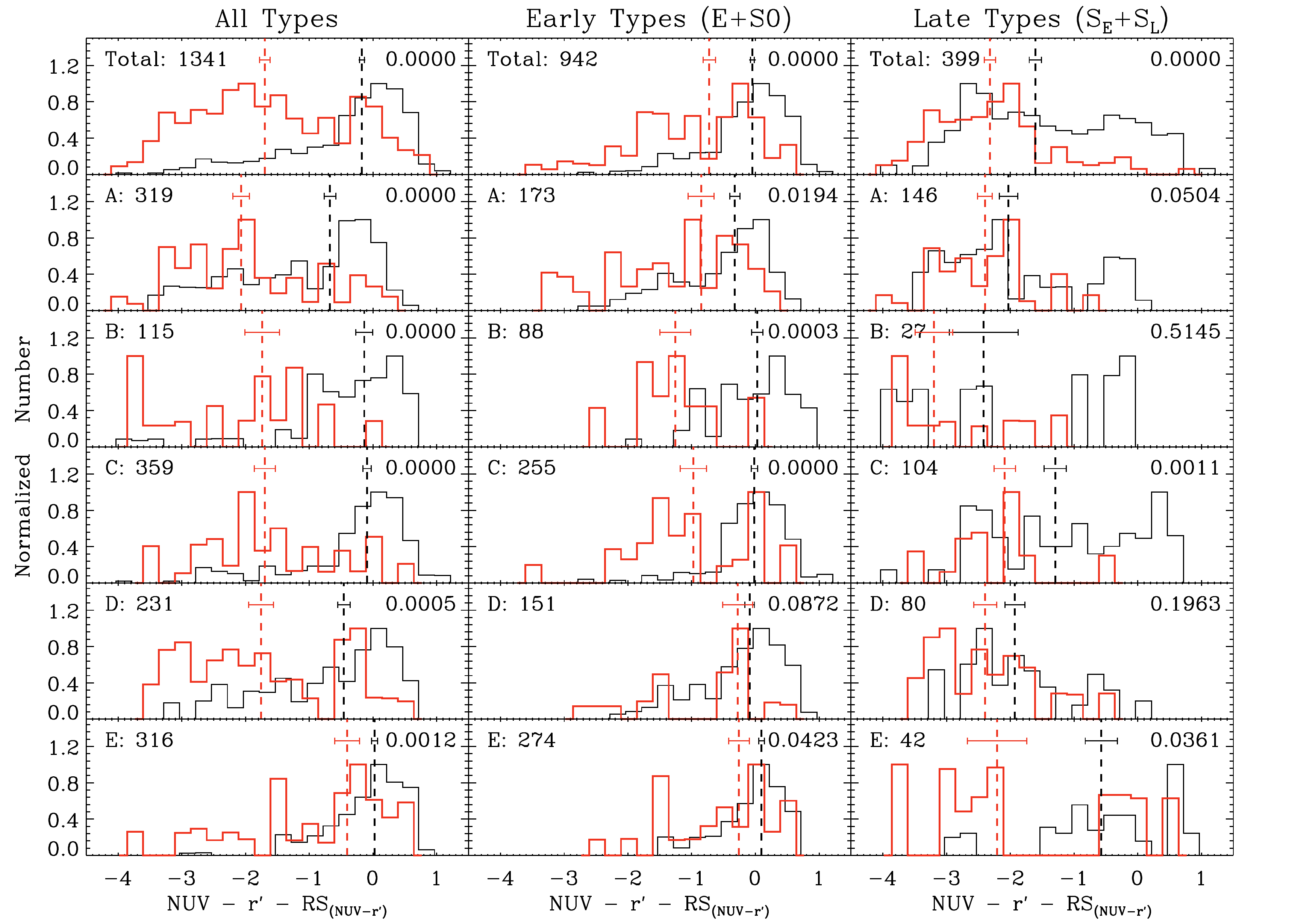}
       \caption[The NUV$-\rmag$ colour distributions]
      {Same as Figure~\ref{phasehist} but with 817 galaxies based on conservative sampling for the NUV band (NUV $<$ 23.5 and $t_{\rm exp}>1500$ seconds).}
     \label{cphasehist}
     \end{figure*}

We present Figure~\ref{cphasehist} which is the same as Figure~\ref{phasehist} only using 817 galaxies with NUV $<$ 23.5 and $t_{\rm exp}>1500$ seconds (dashed line in Figure~\ref{galex}). The overall trends are similar to Figure~\ref{phasehist} although the larger p-values from the KS test indicate lower statistical significances than Figure~\ref{phasehist} for some cases (e.g. Region A for late types and Region D). The correction for the NUV completeness reduces sampling biases based on the NUV detection, hence our results are not affected by sampling.  
  
\clearpage
 \label{lastpage}

\begin{thebibliography}{}

\bibitem[Baldry et al.(2004)]{2004ApJ...600..681B} Baldry, I.~K., Glazebrook, K., Brinkmann, J., et al.\ 2004, \apj, 600, 681 
\bibitem[Blakeslee et al.(2003)]{2003ApJ...596L.143B} Blakeslee, J.~P., Franx, M., Postman, M., et al.\ 2003, \apjl, 596, L143 


\bibitem[Chilingarian et al.(2010)]{2010MNRAS.405.1409C} Chilingarian, I.~V., Melchior, A.-L., \& Zolotukhin, I.~Y.\ 2010, \mnras, 405, 1409 


\bibitem[Choi \& Yi(2017)]{2017ApJ...837...68C} Choi, H., \& Yi, S.~K.\ 2017, \apj, 837, 68 


\bibitem[Cox et al.(2008)]{2008MNRAS.384..386C} Cox, T.~J., Jonsson, P., Somerville, R.~S., Primack, J.~R., \& Dekel, A.\ 2008, \mnras, 384, 386 


\bibitem[Di Matteo et al.(2008)]{2008A&A...492...31D} Di Matteo, P., Bournaud, F., Martig, M., et al.\ 2008, \aaa, 492, 31 


\bibitem[Duc et al.(1997)]{1997A&A...326..537D} Duc, P.-A., Brinks, E., Wink, J.~E., \& Mirabel, I.~F.\ 1997, \aaa, 326, 537 


\bibitem[George(2017)]{2017A&A...598A..45G} George, K.\ 2017, \aaa, 598, A45 


\bibitem[George \& Zingade(2015)]{2015A&A...583A.103G} George, K., \& Zingade, K.\ 2015, \aaa, 583, A103 


\bibitem[Gunn \& Gott(1972)]{1972ApJ...176....1G} Gunn, J.~E., \& Gott, J.~R., III 1972, \apj, 176, 1 


\bibitem[Haines et al.(2015)]{2015ApJ...806..101H} Haines, C.~P., Pereira, M.~J., Smith, G.~P., et al.\ 2015, \apj, 806, 101 


\bibitem[Haines et al.(2015)]{2015MNRAS.451..433H} Haines, T., McIntosh, D.~H., S{\'a}nchez, S.~F., Tremonti, C., \& Rudnick, G.\ 2015, \mnras, 451, 433 


\bibitem[Jeong et al.(2007)]{2007MNRAS.376.1021J} Jeong, H., Bureau, M., Yi, S.~K., Krajnovi{\'c}, D., \& Davies, R.~L.\ 2007, \mnras, 376, 1021 


\bibitem[Jeong et al.(2009)]{2009MNRAS.398.2028J} Jeong, H., Yi, S.~K., Bureau, M., et al.\ 2009, \mnras, 398, 2028 
\bibitem[Juneau et al.(2005)]{2005ApJ...619L.135J} Juneau, S., Glazebrook, K., Crampton, D., et al.\ 2005, \apjl, 619, L135


\bibitem[Jung et al.(2018)]{2018ApJ...865..156J} Jung, S.~L., Choi, H., Wong, O.~I., et al.\ 2018, \apj, 865, 156 
\bibitem[Kauffmann et al.(2003)]{2003MNRAS.341...54K} Kauffmann, G., Heckman, T.~M., White, S.~D.~M., et al.\ 2003, \mnras, 341, 54
\bibitem[Kaviraj(2014)]{2014MNRAS.440.2944K} Kaviraj, S.\ 2014, \mnras, 440, 2944
\bibitem[Kaviraj(2014)]{2014MNRAS.437L..41K} Kaviraj, S.\ 2014, \mnras, 437, L41

\bibitem[Kaviraj et al.(2007)]{2007ApJS..173..619K} Kaviraj, S., Schawinski, K., Devriendt, J.~E.~G., et al.\ 2007, \apjs, 173, 619 


\bibitem[L{\'o}pez-Cruz et al.(2004)]{2004ApJ...614..679L} L{\'o}pez-Cruz, O., Barkhouse, W.~A., \& Yee, H.~K.~C.\ 2004, \apj, 614, 679 


\bibitem[Lawrence et al.(1989)]{1989MNRAS.240..329L} Lawrence, A., Rowan-Robinson, M., Leech, K., Jones, D.~H.~P., \& Wall, J.~V.\ 1989, \mnras, 240, 329 


\bibitem[Lee et al.(2006)]{2006ApJ...650..148L} Lee, J.~H., Lee, M.~G., \& Hwang, H.~S.\ 2006, \apj, 650, 148 


\bibitem[Martin et al.(2005)]{2005ApJ...619L...1M} Martin, D.~C., Fanson, J., Schiminovich, D., et al.\ 2005, \apjl, 619, L1 

\bibitem[Martin et al.(2018)]{2018MNRAS.480.2266M} Martin, G., Kaviraj, S., Devriendt, J.~E.~G., et al.\ 2018, \mnras, 480, 2266

\bibitem[McIntosh et al.(2014)]{2014MNRAS.442..533M} McIntosh, D.~H., Wagner, C., Cooper, A., et al.\ 2014, \mnras, 442, 533 


\bibitem[Mei et al.(2009)]{2009ApJ...690...42M} Mei, S., Holden, B.~P., Blakeslee, J.~P., et al.\ 2009, \apj, 690, 42 


\bibitem[Mei et al.(2006)]{2006ApJ...644..759M} Mei, S., Holden, B.~P., Blakeslee, J.~P., et al.\ 2006, \apj, 644, 759 




\bibitem[Oh et al.(2018)]{2018ApJS..237...14O} Oh, S., Kim, K., Lee, J.~H., et al.\ 2018, \apjs, 237, 14 

\bibitem[Peng et al.(2010)]{2010AJ....139.2097P} Peng, C.~Y., Ho, L.~C., Impey, C.~D., \& Rix, H.-W.\ 2010, \aj, 139, 2097 


\bibitem[Peng et al.(2002)]{2002AJ....124..266P} Peng, C.~Y., Ho, L.~C., Impey, C.~D., \& Rix, H.-W.\ 2002, \aj, 124, 266 

\bibitem[Renzini(2006)]{2006ARA&A..44..141R} Renzini, A.\ 2006, \araa, 44, 141 


\bibitem[Rhee et al.(2017)]{2017ApJ...843..128R} Rhee, J., Smith, R., Choi, H., et al.\ 2017, \apj, 843, 128 

\bibitem[Salim et al.(2009)]{2009ApJ...700..161S} Salim, S., Dickinson, M., Michael Rich, R., et al.\ 2009, \apj, 700, 161 

\bibitem[Sanders et al.(1988)]{1988ApJ...325...74S} Sanders, D.~B., Soifer, B.~T., Elias, J.~H., et al.\ 1988, \apj, 325, 74 


\bibitem[Schawinski et al.(2007)]{2007ApJS..173..512S} Schawinski, K., Kaviraj, S., Khochfar, S., et al.\ 2007, \apjs, 173, 512 


\bibitem[Schawinski et al.(2009)]{2009MNRAS.396..818S} Schawinski, K., Lintott, C., Thomas, D., et al.\ 2009, \mnras, 396, 818 


\bibitem[Schlafly \& Finkbeiner(2011)]{2011ApJ...737..103S} Schlafly, E.~F., \& Finkbeiner, D.~P.\ 2011, \apj, 737, 103 


\bibitem[Sheen et al.(2016)]{2016ApJ...827...32S} Sheen, Y.-K., Yi, S.~K., Ree, C.~H., et al.\ 2016, \apj, 827, 32 


\bibitem[Sheen et al.(2012)]{2012ApJS..202....8S} Sheen, Y.-K., Yi, S.~K., Ree, C.~H., \& Lee, J.\ 2012, \apjs, 202, 8 


\bibitem[Smith et al.(2016)]{2016ApJ...833..109S} Smith, R., Choi, H., Lee, J., et al.\ 2016, \apj, 833, 109 

\bibitem[Smith et al.(2013)]{2013MNRAS.429.1066S} Smith, R., S{\'a}nchez-Janssen, R., Fellhauer, M., et al.\ 2013, \mnras, 429, 1066

\bibitem[Wang et al.(2004)]{2004ApJS..154..193W} Wang, Z., Fazio, G.~G., Ashby, M.~L.~N., et al.\ 2004, \apjs, 154, 193 


\bibitem[Wyder et al.(2007)]{2007ApJS..173..293W} Wyder, T.~K., Martin, D.~C., Schiminovich, D., et al.\ 2007, \apjs, 173, 293 

\bibitem[Yi et al.(2013)]{2013A&A...554A.122Y} Yi, S.~K., Lee, J., Jung, I., Ji, I., \& Sheen, Y.-K.\ 2013, \aaa, 554, A122

\bibitem[Yi et al.(2005)]{2005ApJ...619L.111Y} Yi, S.~K., Yoon, S.-J., Kaviraj, S., et al.\ 2005, \apjl, 619, L111

\bibitem[Zheng et al.(2007)]{2007ApJ...661L..41Z} Zheng, X.~Z., Bell, E.~F., Papovich, C., et al.\ 2007, \apjl, 661, L41


\end{thebibliography}
\end{document}